\documentclass[twocolumn]{aastex62}

\usepackage{ulem}
\hypersetup{linkcolor=red,citecolor=blue,filecolor=cyan,urlcolor=magenta}
\usepackage{amsmath}
\usepackage{tensor}     
\usepackage{graphicx}   
\usepackage{bm}         
\usepackage{xcolor}     
\usepackage{color}      
\usepackage[section]{placeins} 
\usepackage[utf8]{inputenc} 
\newcommand{\nc}{\newcommand*} 
\nc{\figurewidth}{3.2in}
\nc{\xbar}{\bar{x}}
\nc{\rhoeq}{\rho_{\mathrm{eq}}}
\nc{\zeq}{z_{\mathrm{eq}}}
\nc{\tla}{\tilde{\lambda}}
\nc{\dt}{\delta}
\nc{\Dt}{\Delta}
\nc{\vj}{\vec{j}}
\nc{\vl}{\vec{l}}
\nc{\hx}{\hat{x}}
\nc{\hy}{\hat{y}}
\nc{\bj}{\bm{j}}
\nc{\mJ}{\mathcal{J}}
\nc{\mP}{\mathcal{P}}
\nc{\Msun}{M_\odot}
\nc{\app}{\approx}
\nc{\av}[1]{\langle #1 \rangle}
\nc{\eq}[1]{Eq.~\eqref{#1}}
\nc{\al}{\alpha}
\nc{\Xstar}{X_{\ast}}
\nc{\seq}{\sigma_{\mathrm{eq}}}
\nc{\fpbh}{f_{\mathrm{pbh}}}
\nc{\vth}{\vec{\theta}}
\nc{\vla}{\vec{\lambda}}
\nc{\vd}{\vec{d}}
\nc{\Mmin}{M_{\mathrm{min}}}
\nc{\rmd}{\mathrm{d}}
\nc{\mmin}{{m_{\mathrm{min}}}}
\nc{\mmax}{{m_{\mathrm{max}}}}
\nc{\mR}{\mathcal{R}}
\nc{\tmR}{\tilde{\mathcal{R}}}
\nc{\s}{\sigma}
\nc{\ogw}{\Omega_{\mathrm{GW}}}
\nc{\addref}{[\textcolor{red}{add ref}] }
\nc{\Om}{\Omega}
\nc{\gpcyr}{\mathrm{Gpc}^{-3}\,\mathrm{yr}^{-1}}
\nc{\Eq}[1]{Eq.~\eqref{#1}}
\nc{\Fig}[1]{Fig.~\ref{#1}}
\nc{\Table}[1]{Table~\ref{#1}}
\nc{\lvc}{LIGO/Virgo} 
\nc{\Sec}[1]{Sec.~\ref{#1}}
\nc{\eg}{\textit{e.g.~}}
\nc{\SNR}{\mathrm{SNR}}

\def\({\left(}
\def\){\right)}
\def\[{\left[}
\def\]{\right]}

\def\e{\begin{equation}}
\def\q{\end{equation}}
\def\m{\begin{eqnarray}}
\def\n{\end{eqnarray}}
\received{}
\revised{}
\accepted{}
\published{}
\submitjournal{ApJ}
\begin{document}
	
\title{Stochastic Gravitational-Wave Background from Binary Black Holes 
	and Binary Neutron Stars and Implications for LISA}


\author{Zu-Cheng Chen}
\email{chenzucheng@itp.ac.cn} 
\affiliation{CAS Key Laboratory of Theoretical Physics, 
Institute of Theoretical Physics, Chinese Academy of Sciences,
Beijing 100190, China}
\affiliation{School of Physical Sciences, 
University of Chinese Academy of Sciences, 
No. 19A Yuquan Road, Beijing 100049, China}

\author{Fan Huang}
\email{huangfan@itp.ac.cn} 
\affiliation{CAS Key Laboratory of Theoretical Physics, 
Institute of Theoretical Physics, Chinese Academy of Sciences,
Beijing 100190, China}
\affiliation{School of Physical Sciences, 
University of Chinese Academy of Sciences, 
No. 19A Yuquan Road, Beijing 100049, China}

\author{Qing-Guo Huang}
\email{huangqg@itp.ac.cn}
\affiliation{CAS Key Laboratory of Theoretical Physics, 
Institute of Theoretical Physics, Chinese Academy of Sciences,
Beijing 100190, China}
\affiliation{School of Physical Sciences, 
University of Chinese Academy of Sciences, 
No. 19A Yuquan Road, Beijing 100049, China}
\affiliation{Center for Gravitation and cosmology, 
College of Physical Science and Technology, Yangzhou University, 
88 South University Ave., 225009, Yangzhou, China}
\affiliation{Synergetic Innovation Center for Quantum Effects and Applications, 
Hunan Normal University, 36 Lushan Lu, 410081, Changsha, China}

\begin{abstract}
The advent of gravitational wave (GW) and multi-messenger astronomy
has stimulated the research on the formation mechanisms of binary 
black holes (BBHs) observed by LIGO/Virgo. 
In literature, the progenitors of these BBHs could be stellar-origin 
black holes (sBHs) or primordial black holes (PBHs).
In this paper we calculate the Stochastic Gravitational-Wave Background (SGWB) from BBHs,
covering the astrophysical and primordial scenarios separately, 
together with the one from binary neutron stars (BNSs).
Our results indicate that PBHs contribute a stronger SGWB than that from sBHs, 
and the total SGWB from both BBHs and BNSs has a high possibility
to be detected by the future observing runs of  LIGO/Virgo and LISA.
On the other hand, the SGWB from BBHs and BNSs also contributes 
an additional source of confusion noise to LISA's
total noise curve, and then weakens LISA's detection abilities.
For instance, the detection of massive black hole binary (MBHB) coalescences
is one of the key missions of LISA, 
and the largest detectable redshift of MBHB mergers can be significantly reduced.
\end{abstract}

\keywords{gravitational waves -- instrumentation: detectors (LISA) -- 
methods: data analysis -- stars: black holes -- stars: neutron 
}


\section{Introduction}
The detections of gravitational waves (GWs) from binary black hole (BBH)
and binary neutron star (BNS) coalescences by \lvc\ 
\citep{TheLIGOScientific:2016pea,TheLIGOScientific:2016qqj,Abbott:2016nmj,Abbott:2016blz,%
TheLIGOScientific:2017qsa,Abbott:2017vtc,Abbott:2017gyy,Abbott:2017oio}
have led us to the eras of GW and multi-messenger astronomy.
Up to now, there are several BBH merger events reported, of which 
the masses and redshifts are summarized in \Table{events}. 
The progenitors of these BBHs, however, are still under debates.
There exist different formation mechanisms in literature to 
account for the BBHs observed by \lvc.
Under the assumption that all the BBH mergers are of astrophysical origin,
the local merger rate of stellar-mass BBHs is constrained to be 
$12-213\, \gpcyr$ \citep{Abbott:2017vtc}.
Besides, the rate of BNS mergers is estimated to be
$1540\,^{+3200}_{-1220}\, \gpcyr$,
utilizing the only so far observed BNS event, 
GW170817 \citep{TheLIGOScientific:2017qsa}.

Meanwhile, \Table{events} indicates that the masses of BBHs 
extend over a relatively narrow range around $30 \Msun$ with source redshifts 
$z \lesssim 0.2$, due to the detection ability of current generation
of ground-based detectors
(the recent LIGO can measure the redshift of BBH mergers up to $z \sim 1$
\citep{TheLIGOScientific:2016htt,Aasi:2013wya}).
Hence, there are many more unresolved
BBH merger events, along with other sources, emitting energies, 
which can be incoherent superposed to 
constitute a stochastic gravitational-wave background (SGWB)
\citep{Christensen:1992wi}.
Different formation channels for BBHs, in general, predict distinct
mass and redshift distributions for BBH merger rates, 
and thus different energy spectra of SGWBs.
Therefore, the probing of SGWB may serve as a way to discriminate 
various formation mechanisms of BBHs.

Assuming all the black holes (BHs) are of stellar origin
\citep{Belczynski:2010tb,Miller:2016krr,TheLIGOScientific:2016htt,%
Belczynski:2016obo,Stevenson:2017tfq},
the SGWB from BBHs was calculated in
~\cite{TheLIGOScientific:2016wyq,TheLIGOScientific:2016dpb}  
and further updated to include BNSs \citep{Abbott:2017xzg},
indicating that this background would likely to be detectable even 
before reaching \lvc's final design sensitivity, 
in the most optimistic case.
In addition to astrophysical origin, there is another possibility that
the detected BBHs are of primordial origin and (partially) play the role of 
cold dark matter (CDM).
In the early Universe, sufficiently dense regions could undergo 
gravitational collapse by the primordial density inhomogeneity 
and form primordial black holes (PBHs) 
\citep{Hawking:1971ei,Carr:1974nx}.
In literature, two scenarios for PBHs to form  
BBHs exist (see \eg \cite{Garcia-Bellido:2017fdg} and \cite{Sasaki:2018dmp} 
for recent reviews).
The first one is that PBHs in a DM halo interact with each other
through gravitational radiation and occasionally bind to form BBHs
in the late Universe \citep{1989ApJ...343..725Q,Mouri:2002mc,Bird:2016dcv,
Clesse:2016ajp,Clesse:2016vqa}.
The resulting SGWB for the monochromatic mass function is significantly 
lower than that from the stellar origin
and is unlikely to be measured by \lvc~\citep{Mandic:2016lcn},
while the one for a broad mass function could be potentially enhanced
\citep{Clesse:2016ajp}.
The second one is that two nearby PBHs form a BBH due to the tidal
torques from other PBHs in the early Universe
\citep{Nakamura:1997sm,Ioka:1998nz,Sasaki:2016jop}.
The SGWB was investigated in \cite{Wang:2016ana} and \cite{Raidal:2017mfl}, 
showing that it is comparable to that from the stellar-origin BBHs (SOBBHs),
and could serve as a new probe to constrain the fraction of PBHs in CDM.
However, \cite{Raidal:2017mfl} only considered
the tidal torque due to the nearest PBH, 
while \cite{Wang:2016ana} assumed that all the PBHs have the same mass.

\begin{table}[t!]
    \begin{tabular}{c|c|c|c}
        Events &Primary mass &Secondary mass & Redshift\,\\
        \hline
        GW150914\, 
        &  $36.2\,^{+5.2}_{-3.8}$\,$\Msun$ & $29.1\,^{+3.7}_{-4.4}$\,$\Msun$ 
        & $0.09\,^{+0.03}_{-0.04}$\\
        [.3em]
        \hline
        LVT151012\, 
        & $23\,^{+18}_{-6}$\,$\Msun$  & $13\,^{+4}_{-5}$\,$\Msun$ 
        & $0.20\,^{+0.09}_{-0.09}$\\
        [.3em]
        \hline
        GW151226\, 
        &  $14.2\,^{+8.3}_{-3.7}$\,$\Msun$  & $7.5\,^{+2.3}_{-2.3}$\,$\Msun$ 
        & $0.09\,^{+0.03}_{-0.04}$\\
        [.3em]
        \hline
        GW170104\, 
        & $31.2\,^{+8.4}_{-6.0}$\,$\Msun$   & $19.4\,^{+5.3}_{-5.9}$\,$\Msun$  
        & $0.18\,^{+0.08}_{-0.07}$\\
        [.3em]
        \hline
        GW170608\, 
        & $12\,^{+7}_{-2}$\,$\Msun$   & $7\,^{+2}_{-2}$\,$\Msun$  
        & $0.07\,^{+0.03}_{-0.03}$\\
        [.3em]
        \hline
        GW170814\, 
        & $30.5\,^{+5.7}_{-3.0}$\,$\Msun$   & $25.3\,^{+2.8}_{-4.2}$\,$\Msun$  
        & $0.11\,^{+0.03}_{-0.04}$\\
    \end{tabular}
    \caption{A summary of the masses and source redshifts of the six 
        BBHs detected by \lvc\, collaborations
        ~\citep{Abbott:2016blz,TheLIGOScientific:2016pea,TheLIGOScientific:2016qqj,TheLIGOScientific:2016pea,Abbott:2016nmj,Abbott:2017vtc,Abbott:2017gyy,Abbott:2017oio}.
    }
    \label{events}
\end{table}

Recently, the Laser Interferometer Space Antenna (LISA), 
which aims for a much lower frequency regime, 
roughly $10^{-4} \sim 10^{-1}\, \mathrm{Hz}$, than that of \lvc,
has been approved \citep{Audley:2017drz}.
In this paper, we will revisit the SGWB produced by BBHs and BNSs, 
covering both the \lvc\ and LISA frequency band.
The impacts of the SGWB on LISA's detection abilities are also investigated.
For sBHs, we adopt the widely accepted ``Vangioni" model 
\citep{Dvorkin:2016wac} to calculate the corresponding SGWB.
For PBHs, we only consider the early Universe scenario.
The merger rate for PBHs taking into account the torques by all PBHs 
and linear density perturbations was considered in \cite{Ali-Haimoud:2017rtz},
and later improved to encompass the case with a general mass function
for PBHs in \cite{Chen:2018czv}.
We will adopt the merger rate presented in \cite{Chen:2018czv} to estimate
the SGWB from primordial-origin BBHs (POBBHs).
The rest of this paper is organized as follows.
In \Sec{SBH}, assuming all the BBHs are SOBBHs,
we calculate the total SGWB from BBH and BNS mergers following
\cite{Abbott:2017xzg}.
In \Sec{PBH}, assuming all the BBHs are POBBHs and using 
the merger rate density derived in \cite{Chen:2018czv}, 
we estimate the total SGWB from BBHs and BNSs.
Finally, we summarize and discuss our results in \Sec{discuss}

\section{\label{SBH}SGWB from astrophysical binary black holes and binary neutron stars}

There are many different sources in the Universe which can emit 
GWs at different frequency bands. 
Among the various sources, BBHs and BNSs are two of the most 
important ones, which can produce strong SGWB 
and affect LISA's detection abilities.
In this section, we will focus on the SGWB from SOBBHs and BNSs. 

The energy-density spectrum of a GW background can be described by 
the dimensionless quantity \citep{Allen:1997ad}
\e\label{OmegaGW1}
\ogw(\nu) = \frac{\nu}{\rho_{c}} \frac{\text{d}\rho_{\mathrm{GW}}}{\rmd \nu},
\q
where $\rmd \rho_{\mathrm{GW}}$ is the energy density in the frequency 
interval $\nu$ to $\nu+\rmd \nu$, $\rho_{c}=3H_{0}^2 c^2 /(8 \pi G)$ is the 
critical energy density of the Universe, 
and $H_0 = 67.74\, \mathrm{km}\, \mathrm{s}^{-1} \,\mathrm{Mpc}^{-1}$ 
is the Hubble constant taken from Planck \citep{Ade:2015xua}.
For the binary mergers, the magnitude of a SGWB can be further transformed to
\citep{Phinney:2001di,Regimbau:2008nj,Zhu:2011bd,Zhu:2012xw}
\e\label{OmegaGW}
\begin{split}
	\ogw(\nu) =& \frac{\nu}{\rho_c H_0} \int_0^{z_{\mathrm{max}}} \rmd z \int \rmd m_{1} \rmd m_{2}\\
	& \frac{\mR (z,m_{1},m_{2}) \frac{\rmd E_{\mathrm{GW}}}{\rmd \nu_s}(\nu_s,m_{1},m_{2})}{(1+z)E(\Om_r, \Om_m,\Om_{\Lambda},z)},
\end{split}
\q
where $\nu_s = (1+z) \nu$ is the frequency in source frame, 
and $E(\Om_r, \Om_m, \Om_{\Lambda}, z)=\sqrt{\Om_r \(1+z\)^4 + \Om_m (1+z)^3+\Omega_{\Lambda}}$ 
accounts for the dependece of comoving volume on redshift $z$. 
We adopt the best-fit results from Planck \citep{Ade:2015xua} that 
$\Om_r = 9.15 \times 10^{-5}$, $\Om_m = 0.3089$, 
and $\Om_\Lambda = 1 - \Om_m - \Om_r$.
For the cut-off redshift $z_{\mathrm{max}}$, we choose $z_{\mathrm{max}}=10$ for SOBBHs
\citep{TheLIGOScientific:2016wyq},
and $z_{\mathrm{max}} = \nu_3/\nu - 1$ for POBBHs \citep{Wang:2016ana},
in which $\nu_3$ is given by \Eq{dEdnu} below.
The energy spectrum emitted by a single BBH  $\rmd E_{\mathrm{GW}}/\rmd \nu_{s}$
is well approximated by
\citep{Cutler:1993vq,Chernoff:1993th,Zhu:2011bd}
\e\label{dEdnu} 
\hspace{-5mm}\frac{\rmd E_{\mathrm{GW}}}{\rmd \nu_s} = \frac{\(\pi G\)^{2/3} M^{5/3} \eta}{3} \begin{cases}
	\nu_s^{-1/3}, &\hspace{-2mm} \nu_s<\nu_1,\\
	\frac{\nu_s}{\nu_1} \nu^{-1/3}, &\hspace{-2mm} \nu_1 \leq \nu_s < \nu_2,\\
	\frac{\nu_s^2}{\nu_1 \nu_2^{4/3}} \frac{\nu_4^4}{\(4\(\nu_s-\nu_2\)^2 + \nu_4^2\)^2}, 
	&\hspace{-2mm} \nu_2 \leq \nu_s < \nu_3,
\end{cases}
\q
where $\nu_i = \(a_i \eta^2 + b_i \eta + c_i\)/\(\pi G M/ c^3 \)$,
$M = m_1 + m_2$ is the total mass of the binary, 
and $\eta = m_1 m_2 / M^2$. 
The coefficients $a_i$, $b_i$ and $c_i$ can be found in Table~I of \cite{Ajith:2007kx}. 
Since the frequency band of non-zero eccentricity during inspiral phase 
is below $10^{-4}$ Hz \citep{Dvorkin:2016wac}, 
which is beyond the frequency range of LISA,
we hence only consider the circular orbit during inspiral phase. 
A careful discussion of the impact of eccentricity on the SGWB can be found in \cite{DOrazio:2018jnv}. 

We will follow the widely accepted ``Vangioni" model \citep{Dvorkin:2016wac} 
to estimate the SGWB from SOBBHs and BNSs in the Universe.
The merger rate density $\mR(z,m_{1},m_{2})$ in \Eq{OmegaGW}
for the SOBBHs or BNSs is a convolution of the sBHs or neutron stars (NSs)
formation rate
$R_{\mathrm{birth}}(z,m_{1})$ with the distribution of the time delays 
$P_{d} \left(t_{d} \right)$ between the formation and merger
of SOBBHs or BNSs,
\e\label{sBHR}
    \mR= N \int^{t_{\mathrm{max}}}_{t_{\text{min}}}  
        \frac{R_{\mathrm{birth}} (t(z)-t_d,m_1 )\,\times P_d (t_d)}
            {\min(m_1, \mmax-m_1) - \mmin}\ \rmd t_d,
\q
where $N$ is a normalization constant, 
$t(z)$ is the age of the Universe at merger. 
Here $P_{d} \propto t_{d}^{-1}$ is the distribution of delay time $t_{d}$
with $t_{\mathrm{min}} < t_{d} < t_{\mathrm{max}}$ \citep{Abbott:2017xzg}.
The minimum delay time of a massive binary system to evolve until 
coalescence are set to $t_{\mathrm{min}} = 50$\,Myr for SOBBHs,
and $t_{\mathrm{min}} = 20$\,Myr for BNSs.
Meanwhile, the maximum delay time $t_{\mathrm{max}}$ is set to the Hubble time. 
In order to comply with the previous studies \citep{Abbott:2017vtc,Abbott:2017xzg}, 
we restrict the component masses of BBHs to the range 
$\mmin \leq m_2 \leq m_1$
and $m_1 + m_2 \leq \mmax$, with $ \mmin = 5\Msun$ and $\mmax = 100\Msun$.
We note that the merger rate density of POBBHs (see \Eq{calR} below) is quite
different from \Eq{sBHR}, due to the distinct formation mechanisms of
POBBHs and SOBBHs.

The most complicated part of \Eq{sBHR} is the computation of the birthrate 
of sBHs or NSs, which is given by \citep{Dvorkin:2016wac}
\e
\begin{split}
R_{\mathrm{birth}}(t,m_{\mathrm{rem}})=&\int \psi [t-\tau(m_{*})] \phi(m_{*}) \\
 &\delta(m_{*}- g_{\mathrm{rem}}^{-1}(m_{\mathrm{rem}})) \text{d}m_{*} \label{Rbirth},
\end{split}
\q
where $m_{*}$ is the mass of the progenitor star, 
$m_{\mathrm{rem}}$ is the mass of remnant,
and $\tau(m_{*})$ is the lifetime of a progenitor star, 
which can be ignored \citep{Schaerer:2001jc}.
Here $\phi(m_{*})$ is the so called initial mass function (IMF),
which is a uniform distribution ranging from $1\,\Msun$ to $2\,\Msun$ for NSs
and $\phi(m_{*}) \propto m_{*}^{-2.35}$ for sBHs.
In addition, $\psi(t)$ is the star formation rate (SFR), 
which is given by \citep{Nagamine:2003bd}
\e
\psi(z)= k \frac{a \exp[b(z-z_{m})]}{a-b+b \exp[a(z-z_{m})]}. 
\q
We will use the fit parameters given by ``\textit{Fiducial+PopIII}" model 
from \cite{Dvorkin:2016wac}, namely, the sum of \textit{Fiducial} SFR 
(with $k=0.178 \, \Msun \, \mathrm{yr}^{-1} 
\, \mathrm{Mpc}^{-3}$, $z_{m}=2$, $a=2.37$, $b=1.8$) 
and \textit{PopIII} SFR (with $k=0.002 \, \Msun 
\, \mathrm{yr}^{-1} \, \mathrm{Mpc}^{-3}$, $z_{m}=11.87$, $a=13.8$ , $b=13.36$).
Dirac delta function in \Eq{Rbirth} relates to the process of BH formation. 
For NSs, $g^{-1}_{\mathrm{ns}}(m_{\mathrm{ns}})=m_{\mathrm{ns}}$, 
one obtains a relative simple form of birthrate. 
However, for sBHs, the masses of the progenitor star and the remnant 
are related by some function $m_{\mathrm{bh}}=g_{\mathrm{bh}}(m_{*})$,
which is model-dependent and still unclear yet. 
In this paper we consider the \textit{WWp} model 
\citep{Woosley:1995ip} of sBH formation,
which is simple and indistinguishable from the widely used Fryer model
at low redshift \citep{Dvorkin:2016wac}.
For progenitor with initial mass $m_{*}$, the mass of the remnant BH
$m_{\mathrm{bh}}$ is extrapolated as
\e
\frac{m_{\mathrm{bh}}}{m_{*}}=A \left(\frac{m_{*}}{40\Msun} \right)^{\beta} \frac{1}{\left( \frac{Z(z)}{0.01 Z_\odot} \right)^{\gamma} +1},
\q
where $Z(z)$ is the metallicity and an explicit functional 
form can be found in \cite{Belczynski:2016obo}. 
The fiducial values of this extrapolation are $A=0.3$, $\beta =0.8$ 
and $\gamma =0.2$ \citep{Dvorkin:2016wac}.
Solving the equation above yields the function
$m_{*} = g_{\mathrm{bh}}^{-1}(m_{\mathrm{bh}})$.

\begin{figure}[htbp!]
	\centering
	\includegraphics[width = 0.48\textwidth]{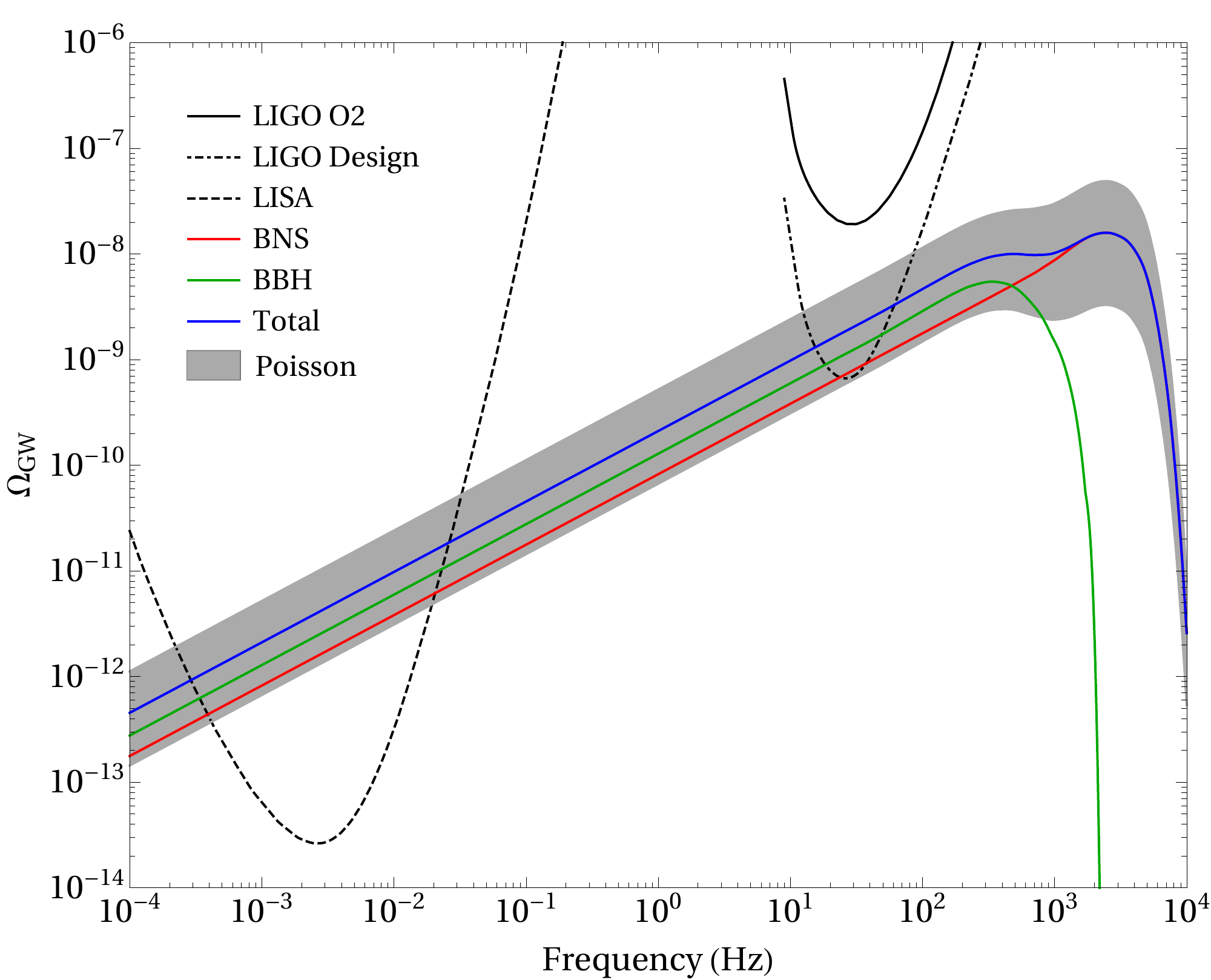}
	\caption{\label{OmegaGW-sBH}
    The predicted SGWB from the BNSs and SOBBHs. 
    The red and green curves are backgrounds from the BNSs and BBHs,
    respectively. 
    The total (BNS and BBH) background is shown in the blue curve, while
    its Poisson error bars are in the grey shaded region.
    Here, we adopt the local merger rate $R = 103_{-63}^{+110}$\,$\gpcyr$ 
    for SOBBHs \citep{Abbott:2017vtc},
    and $R = 1540_{-1220}^{+3200}$\,$\gpcyr$ for BNSs
    \citep{TheLIGOScientific:2017qsa}.
    We also show the expected PI curves for LISA with $4$ years of 
    observation (dashed) and
    LIGO's observing runs of O2 (black) and design sensitivity (dot-dashed).
    The PI curves for LISA and LIGO's design sensitivity cross the Poisson
    error region, indicating the possibility to detect this background.
    }
\end{figure}

Integrating over the component masses in merger rate density, 
results in the merger rate as a function of redshift  
\e
 \mR(z)  =\int \mR(z,m_{1},m_{2} )\, \rmd m_{1} \text{d}m_{2}.
\q
The local merger rate $R \equiv \mR(z=0)$ is inferred to be
$R = 103_{-63}^{+110}$\,$\gpcyr$ for SOBBHs \citep{Abbott:2017vtc},
and $R = 1540_{-1220}^{+3200}$\,$\gpcyr$ for BNSs
\citep{TheLIGOScientific:2017qsa}.
Utilizing \Eq{OmegaGW}, we then calculate the SGWB from SOBBHs and BNSs.
In \Fig{OmegaGW-sBH}, we show the corresponding SGWBs as well as
the power-law integrated (PI) curves of LIGO \citep{TheLIGOScientific:2017qsa} 
and LISA \citep{Cornish:2017vip,Cornish:2018dyw},
indicating that the total SGWB from both BBHs and BNSs has a high possibility
to be detected by the future observing runs of \lvc\, and LISA.
The energy spectra from both the SOBBHs and BNSs are well approximated by 
$\ogw \propto \nu^{2/3}$ at low frequencies covering both
LISA and LIGO's bands, 
where the dominant contribution is from the inspiral phase.
We also summarize the background energy densities $\ogw (\nu)$ 
at the most sensitive frequencies of LIGO (near $25$ Hz) and LISA 
(near $3\times 10^{-3}$ Hz) in \Table{Omegaf-sBH}.

\begin{table}[htbp!]
\begin{tabular}{c|c|c}
    &\ $\Omega_{\mathrm{GW}}(25 \, \mathrm{Hz})$ \
    &\ $\Omega_{\mathrm{GW}}(3 \times 10^{-3} \, \mathrm{Hz})$\,\\
	\hline
	BNS\, &  $0.7^{+1.5}_{-0.6} \times 10^{-9}$ 
          & $1.7^{+3.5}_{-1.4} \times 10^{-12}$ \\
	[.3em]
	\hline
	BBH\, & $1.1^{+1.2}_{-0.7} \times 10^{-9}$  
          & $2.7^{+2.8}_{-1.6} \times 10^{-12}$ \\
	[.3em]
	\hline
	Total\, & $1.8^{+2.7}_{-1.3} \times 10^{-9}$  
            & $4.4^{+6.3}_{-3.0} \times 10^{-12}$ \\
	[.2em]
\end{tabular}
    \caption{\label{Omegaf-sBH}
    Estimates of the background energy density $\ogw (\nu)$ at the most
    sensitive frequencies of LIGO (near $25$ Hz) and LISA 
    (near $3\times 10^{-3}$ Hz) for each of the BNS, SOBBH and 
    total background contributions, 
    along with the $90\%$ Poisson error bounds.
	}
\end{table}

\begin{figure}[htbp!]
	\centering
	\includegraphics[width = 0.48\textwidth]{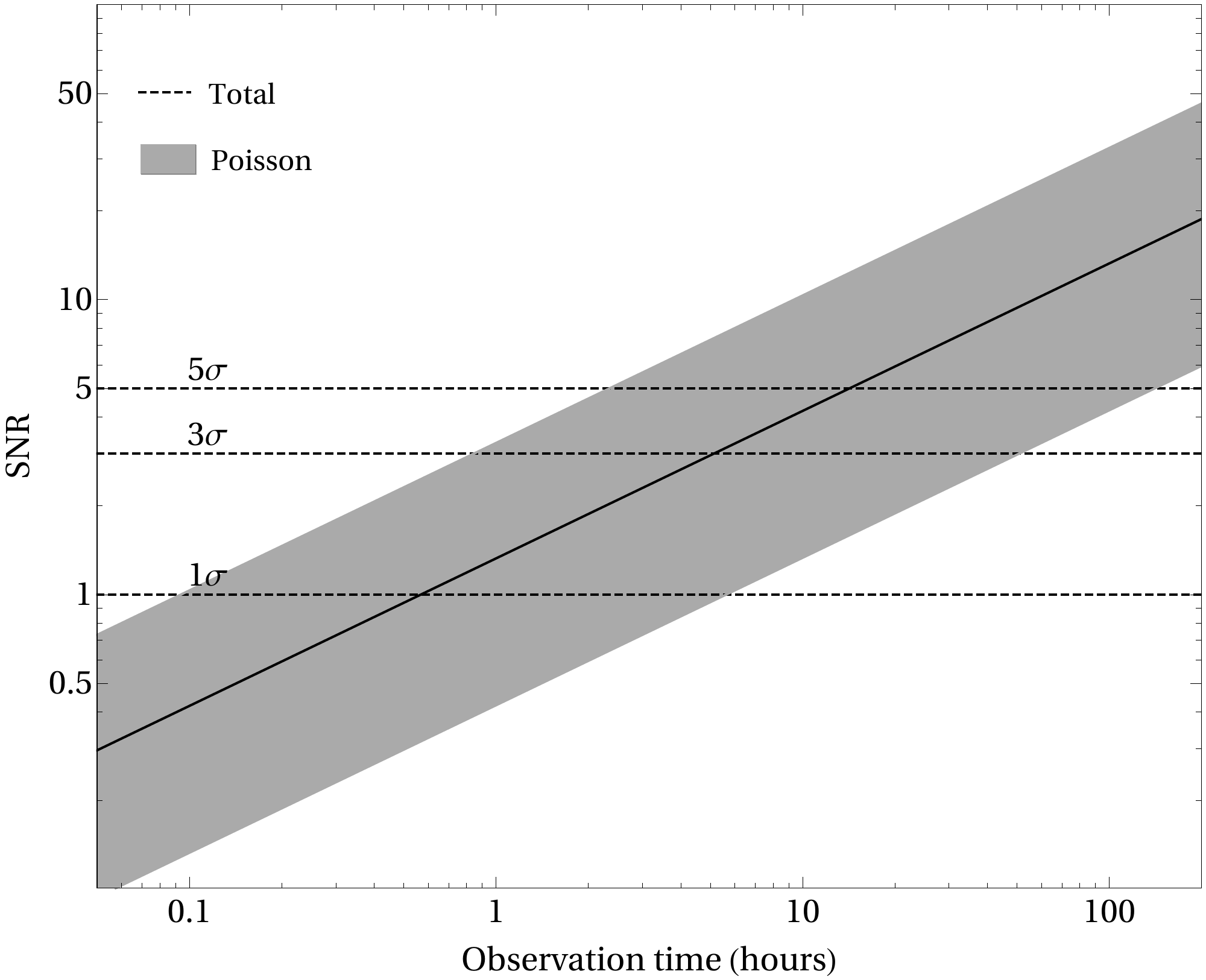}
	\caption{\label{SNR-sBH} 
    The SNR of LISA as a function of observing time for median total 
    SGWB (black curve) and associated uncertainties (grey shaded region),
    from the SOBBHs and BNSs.
    Here, we adopt the local merger rate $R = 103_{-63}^{+110}$\,$\gpcyr$ 
    for SOBBHs \citep{Abbott:2017vtc},
    and $R = 1540_{-1220}^{+3200}$\,$\gpcyr$ for BNSs
    \citep{TheLIGOScientific:2017qsa}.
    The predicted median total background can be detected with 
    $\mathrm{SNR}=5$ after about $20$ hours of observation time.
    }
\end{figure}

\begin{figure}[htbp!]
	\centering
	\includegraphics[width = 0.48\textwidth]{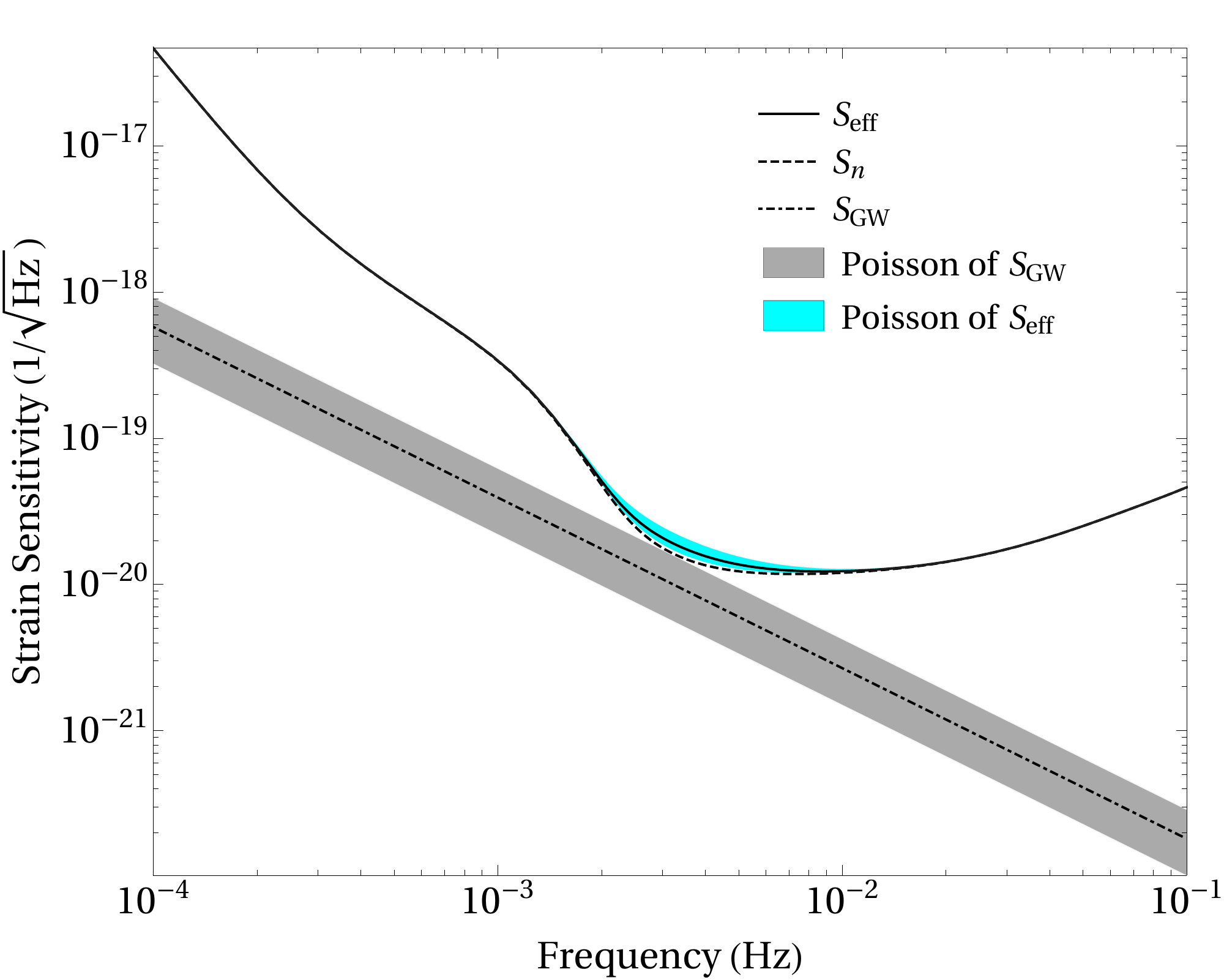}
	\caption{\label{lisa-sensitivity-sBH}
		The effective strain sensitivity $S_{\mathrm{eff}}$ (black solid curve) of LISA
        and its Poisson uncertainties (cyan region),
        due to the effect of the total SGWB from SOBBHs and BNSs.
        We also show LISA's strain sensitivity $S_n$ (dashed curve),
        and $S_{\mathrm{GW}}$ (dot-dashed curve) 
        along with its Poisson uncertainties (grey shaded region).
    }
\end{figure}

\begin{figure}[htbp!]
	\centering
	\includegraphics[width = 0.48\textwidth]{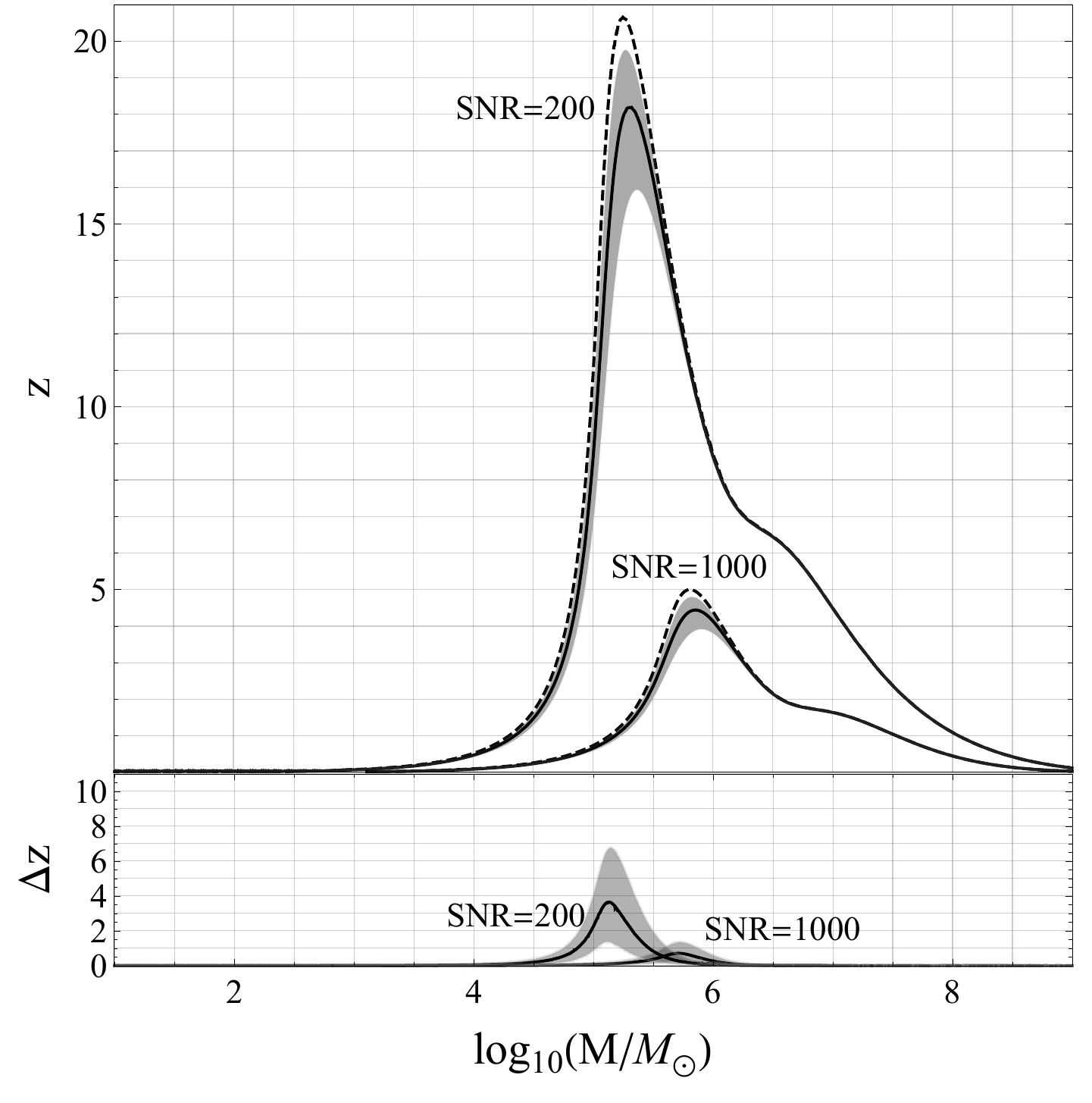}
	\caption{\label{z-sBH}
        The impacts of the total SGWB from SOBBHs and BNSs 
        on the largest detectable redshift $z$ of MBHB (with total mass $M$) coalescences for LISA.
        The mass ratio is set to $q=0.2$ following \cite{Audley:2017drz}. 
        The upper panel shows the contours of $\SNR=200$ and $\SNR=1000$ 
        for LISA (dashed curves), 
        together with the effect of SGWB (black curves) and the Poisson error 
        bars (grey shaded region). 
        The lower panel shows the residuals of corresponding contours. 
    }
\end{figure}
The signal-to-noise ratio (SNR) for measuring the SGWB of LISA with 
observing time $T$ is given by \citep{Thrane:2013oya,Caprini:2015zlo}
\e
\text{SNR}=\sqrt{T} \left[ \int \rmd \nu \frac{\Omega_{\mathrm{GW}}(\nu)}{\Omega_{n}(\nu)} \right] ^{1/2},
\q
where $\Omega_{n}(\nu)\equiv 2 \pi^2 \nu^3 S_{n}(\nu)/\left(3 H_{0}^2 \right)$
and $S_{n}$ is the sensitivity of LISA. 
\Fig{SNR-sBH} shows the expected accumulated SNR of LISA as a 
function of observation time.
The predicted median total background from BBHs and BNSs may be 
identified with $\mathrm{SNR} = 5$ after about $20$ hours of observation.
The total background could be identified with $\mathrm{SNR} = 5$ 
within $3$ hours of observation for the most optimistic case,  
and after about $8$ days for the most pessimistic case.

The total SGWB due to SOBBHs and BNSs is so strong that it may 
become an unresolved noise, 
affecting the on-going missions of LISA.
For instance, the detection of massive black hole binary (MBHB) coalescences
is one of the key missions of LISA \citep{Audley:2017drz}, 
and the largest detectable redshift of a MBHB merger may be 
significantly reduced by the additional noise.

Following \cite{Barack:2004wc} and \cite{Cornish:2018dyw}, 
we define the noise strain sensitivity due to the SGWB as
\e
    S_{\mathrm{GW}}(\nu) \equiv \frac{3 H_{0}^2}{2 \pi^2}
        \frac{\Omega_{\mathrm{GW}}(\nu)}{\nu^3},
\q
which can be added to the strain sensitivity of LISA $S_{n}(\nu)$
to obtain an effective full strain sensitivity $S_{\mathrm{eff}}(\nu)=S_{n}(\nu)+S_{\mathrm{GW}}(\nu)$. 
The resulting strain sensitivity curves are shown in \Fig{lisa-sensitivity-sBH}.
Additionally, the SNR of an single incoming GW strain signal
(also called waveform) $h(t)$ has the following form 
\e
    \SNR = 2 \left[ \int \rmd \nu 
        \frac{|\tilde{h}(\nu)|^2}{S_{\mathrm{eff}}(\nu)} \right]^{1/2},
\q
where $\tilde{h}(\nu)$ is the frequency domain representation of $h(t)$,
and we adopt the phenomenological waveform provided by \cite{Ajith:2007kx}.

Study the growth mechanism of massive black holes (MBHs) is an important 
science investigation (SI) of LISA \citep{Audley:2017drz}.
Among the observational requirements of that SI, 
being able to measure the dimensionless spin of the largest MBH 
with an absolute error less than 0.1 and detect the misalignment of 
spins with the orbital angular momentum better than $10^{\circ}$,
requires an accumulated SNR of at least 200.
The effect on SNR of MBHB coalescences due to unsolved SGWB signal 
is shown in \Fig{z-sBH}, 
indicating that the largest detectable redshift 
(with a fixed $\SNR=200$ or $\SNR=1000$) 
will be reduced.
It means that the total detectable region of LISA is suppressed, 
thus decreasing the event rate of LISA's scientific missions.
Currently, the studies of the origin of MBHs predict that masses of the 
seeds of MBHs lie in the range about $10^3 \Msun$ to several $10^5 \Msun$, 
with formation redshift around $10\lesssim z \lesssim 15$
\citep{Volonteri:2010wz}.
As shown in \Fig{z-sBH}, the precise measurement of those seeds 
above $10^5 \Msun$ in high formation redshift will be significantly affected 
by the confusion noise of the unsolved SGWB. 
Therefore, further analysis is needed to subtract the SGWB signals
from the data in order to improve the performance of the detectors.

\section{\label{PBH}SGWB from binary primordial black holes}

In this section, we will calculate the SGWB from PBHs assuming all 
BHs observed by \lvc\ so far are of primordial origin. 
Here, we adopt the merger rate for POBBHs presented in \cite{Chen:2018czv},
which takes into account the torques both by all PBHs 
and linear density perturbations.
For a general normalized mass function with parameters $\vth$, 
or the probability distribution function (PDF) for PBHs $P(m|\vth)$, 
the comoving merger rate density in units of $\gpcyr$ is given by
\citep{Chen:2018czv}
\m\label{calR} 
\mR_{12}&&(t|\vth) \approx 3.9\cdot 10^6\times \({t\over t_0}\)^{-{34\over 37}} f^2 (f^2+\sigma_{\mathrm{eq}}^2)^{-{21\over 74}} \nonumber \\
&& \times  \min\(\frac{P(m_1|\vth)}{m_1}, \frac{P(m_2|\vth)}{m_2}\) \({P(m_1|\vth)\over m_1}+{P(m_2|\vth)\over m_2}\) \nonumber \\
&& \times (m_1 m_2)^{{3\over 37}} (m_1+m_2)^{36\over 37},
\n
where $t_0$ is the age of our Universe, and
$\sigma_{\mathrm{eq}}$ is the variance of density perturbations of the rest
DM on scale of order ${\cal O}(10^0\sim10^3) M_\odot$ 
at radiation-matter equality. 
The component masses of a POBBH, $m_1$ and $m_2$, are in units of $\Msun$.
Similar to \cite{Ali-Haimoud:2017rtz} and \cite{Chen:2018czv}, we take $\sigma_{\mathrm{eq}}\approx 0.005$. 
Here $f$ is the total abundance of PBHs in non-relativistic matter,
and the fraction of PBHs in CDM is related to $f$ by $\fpbh\equiv \Omega_{\mathrm{pbh}}/\Omega_{\mathrm{cdm}} \approx f/0.85$. 
Integrating over the component masses, yields the merger rate 
\e
    \mR(t|\vth) = \int \mR_{12}(t|\vth)\ \rmd m_1\, \rmd m_2,
\q 
which is time (or redshift) dependent.
The local merger rate density distribution then follows
\e 
    \mR_{12}(t_0|\vth) = R\, p(m_1,m_2|\vth),
\q 
where $p(m_1,m_2|\vth)$ is the distribution of BH masses in coalescing
binaries. 
The local merger rate $R \equiv \mR(t_0|\vth)$ is a normalization
constant, such that the population distribution $p(m_1,m_2|\vth)$
is normalized. 
Note that all masses are source-frame masses.

We are then interested in extracting the population parameters
$\{\vth, R\}$ from the merger events observed by \lvc.
This is accomplished by performing the hierarchical Bayesian inference
on the BBH's mass distribution 
\citep{Abbott:2016nhf,Abbott:2016drs,TheLIGOScientific:2016pea,
Wysocki:2018mpo,Fishbach:2018edt,Mandel:2018mve,Thrane:2018qnx}.
Given the data for $N$ detections, $\vd = (d_1, \dots, 
d_N)$, the likelihood for an inhomogeneous Poisson process, reads
\citep{Wysocki:2018mpo,Fishbach:2018edt,Mandel:2018mve,Thrane:2018qnx}
\e\label{likelihood}
    p(\vd|\vth, R) \propto R^{N} e^{-R\, \beta(\vth)} \prod_i^N 
            \int \rmd\vla\ p(d_i|\vla)\ p(\vla|\vth),
\q 
where $\vla \equiv \{m_1, m_2\}$, and $p(d_i|\vla)$ is the likelihood of an
individual event with data $d_i$ given the binary parameters $\vla$.
Since the standard priors on masses 
for each event in \lvc\ analysis are taken to be uniform, 
one has $p(d_i|\vla) \propto p(\vla|d_i)$, 
and we can use the announced posterior samples
\citep{Vallisneri:2014vxa,TheLIGOScientific:2016pea,Biwer:2018osg}
to evaluate the integral in \Eq{likelihood}.
Meanwhile, $\beta(\vth)$ is defined as
\e 
    \beta(\vth) \equiv \int \rmd\vla\ VT(\vla)\ p(\vla|\vth),
\q 
where $VT(\vla)$ is the sensitive spacetime volume of LIGO.
We adopt the semi-analytical approximation from 
\cite{Abbott:2016nhf,Abbott:2016drs} 
to calculate $VT$. 
Specifically, we neglect the effect of spins for BHs, and use
aLIGO ``Early High Sensitivity" scenario to approximate
the power spectral density (PSD) curve.
We also consider a single-detector SNR threshold $\rho_{\mathrm{th}} = 8$ for
detection, which is roughly corresponding to a network threshold of $12$.

The posterior probability function $p(\vth, R|\vd)$ of the 
population parameters $\{\vth, R\}$ can be computed by using some 
assumed prior $p(\vth, R)$,
\e\label{post} 
    p(\vth, R|\vd) \propto p(\vd |\vth, R)\ p(\vth, R).
\q 
We take uniform priors for $\vth$ parameters, and a log-uniform
one for local merger rate $R$, thus having
\e 
    p(\vth, R) \propto \frac{1}{R}.
\q 
With this prior in hand, the posterior marginalized over $R$ 
could be easily obtained
\e\label{post_vth} 
    p(\vth|\vd) \propto \[\beta(\vth)\]^{-N} 
               \prod_i^N \int \rmd\vla\ p(d_i|\vla)\ p(\vla|\vth).
\q 
This posterior has been used in previous population inferences 
\citep{Abbott:2016nhf,Abbott:2017vtc,TheLIGOScientific:2016pea,
Abbott:2016drs,Fishbach:2017zga}.
We will follow the same procedure as in \cite{Abbott:2016nhf,
TheLIGOScientific:2016pea,Abbott:2016drs,Abbott:2017vtc}, 
by first using \Eq{post_vth} to constrain
the parameters $\vth$, and then fixing $\vth$ to their best-fit values
in \Eq{post} to infer the local merger rate $R$.
As done in the \Sec{SBH}, 
we also restrict the component
masses of BBHs to the range $5\Msun \leq m_2 \leq m_1$
and $m_1 + m_2 \leq 100\Msun$.
At the time of writing, data analysis for LIGO's O2 observing run
is still on going, we therefore only use the $3$ events from LIGO's 
O1 observing run, which contains $48.6$ days of observing time
\citep{TheLIGOScientific:2016pea}.
An update analysis could be performed until the final release of 
LIGO's O2 samples.
In the following subsections, we will consider two distinct mass functions
for PBHs. We will firstly constrain the population parameters $\{\vth, R\}$
using LIGO's O1 events, and then calculate the corresponding SGWB 
from the inferred results.

\subsection{Power-law mass function}
We now consider a power-law PBH mass functions \citep{Carr:1975qj}
\e\label{power}
 P(m)\approx {\alpha-1\over \Mmin} \({m\over \Mmin}\)^{-\alpha},
\q
for $m\geq \Mmin = 5\Msun$ and $\alpha>1$ is the power-law slope.
In this case, the free parameters are $\{\vth, R\} = \{\al, R\}$.

\begin{figure}[htbp!]
\centering
\includegraphics[width = 0.48\textwidth]{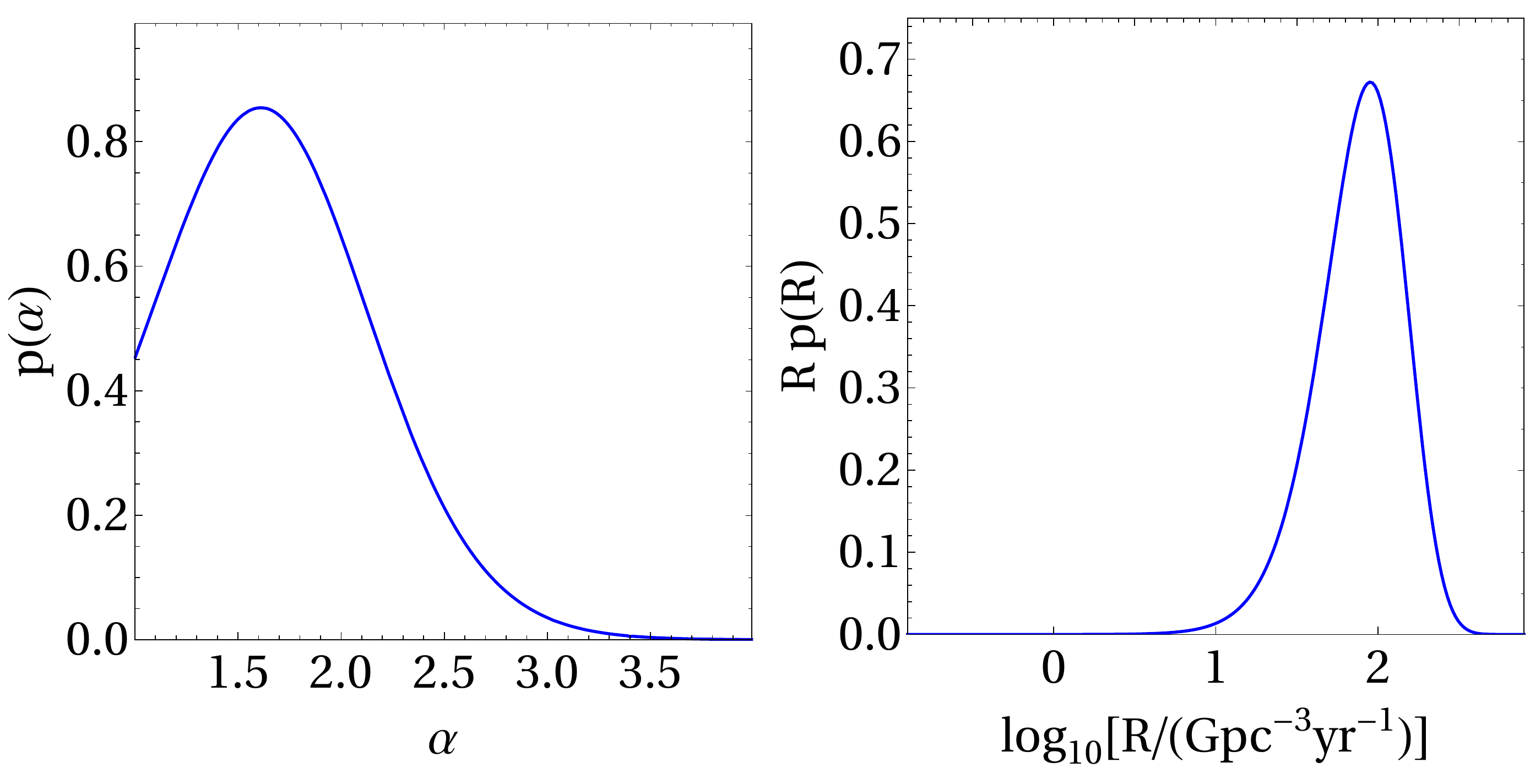}
\caption{\label{posterior-PBH-power} 
    The posterior distributions for $\{\vth, R\} = \{\al, R\}$ for
    \textit{power-law} mass function of PBHs, by using $3$ events from 
    LIGO's O1 observing run.
    }
\end{figure}

Accounting for the selection effects and using $3$ events from LIGO's
O1 run, we find the best-fit result for $\al$ is $\al = 1.61$.
Fixing $\al$ to this best-fit value,
we obtain the median value and $90\%$ equal tailed credible
interval for the local merger rate,
$R = 80\,^{+108}_{-56}$\, $\gpcyr$.
The posterior distributions are shown in \Fig{posterior-PBH-power}.
From the posterior distribution of local merger rate $R$,
we then infer the fraction of PBHs in CDM 
$\fpbh = 3.8\,^{+2.3}_{-1.8} \times 10^{-3}$. 
Such an abundance of PBHs is consistent with previous estimations that
$10^{-3} \lesssim \fpbh \lesssim 10^{-2}$,
confirming that the dominant fraction of CDM should not 
originate from the stellar mass PBHs
\citep{Sasaki:2016jop,Ali-Haimoud:2017rtz,Raidal:2017mfl,%
Kocsis:2017yty,Chen:2018czv}. 

\begin{figure}[htbp!]
\centering
\includegraphics[width = 0.48\textwidth]{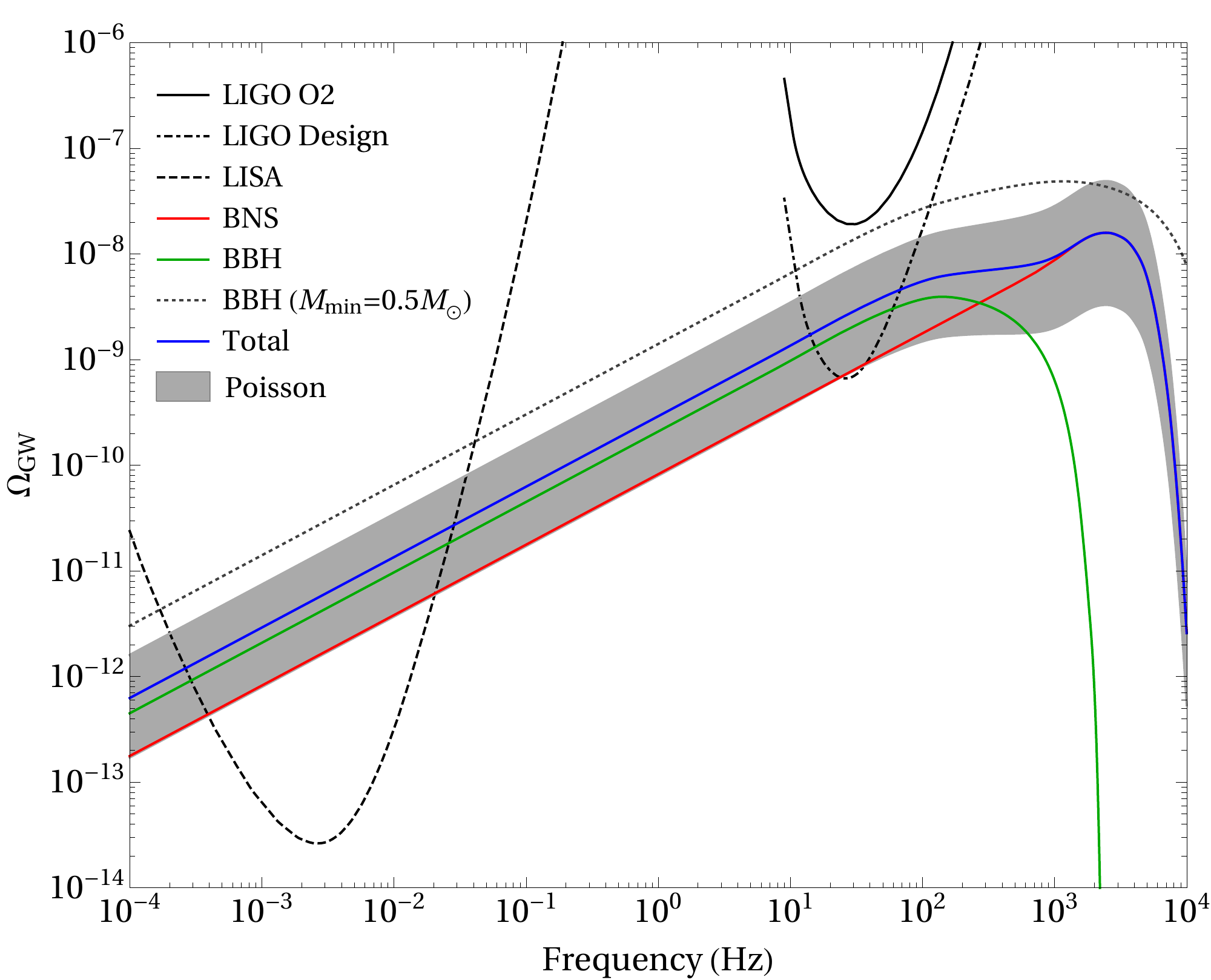}
\caption{\label{OmegaGW-PBH-power} 
    The predicted SGWB from the BNSs and POBBHs
    with a \textit{power-law} mass function. 
    The red and green curves are backgrounds from the BNSs and BBHs, respectively. 
    The total (BNS and BBH) background is shown in the blue curve, while
    its Poisson error bars are in the grey shaded region.
    For BBHs, we adopt the best-fit value for $\al = 1.61$, 
    and the inferred local merger rate $R = 80\,^{+108}_{-56}$\, $\gpcyr$,
    which corresponds to $\fpbh = 3.8\,^{+2.3}_{-1.8} \times 10^{-3}$. 
    And for BNSs, we adopt $R = 1540_{-1220}^{+3200}$\,$\gpcyr$ 
    \citep{TheLIGOScientific:2017qsa}.
    The dotted line shows the background from BBHs with $\Mmin = 0.5 \Msun$, 
    by fixing $\al = 1.61$ and $R = 80$\, $\gpcyr$.
    We also show the expected PI curves for LISA with $4$ years of 
    observation (dashed) and
    LIGO's observing runs of O2 (black) and design sensitivity (dot-dashed).
    The PI curves for LISA and LIGO's design sensitivity cross the Poisson
    error region, indicating the possibility to detect this background or
    set upper limits on the population parameters $\{\al, R\}$.
    }
\end{figure}

\begin{figure}[htbp!]
\centering
\includegraphics[width = 0.48\textwidth]{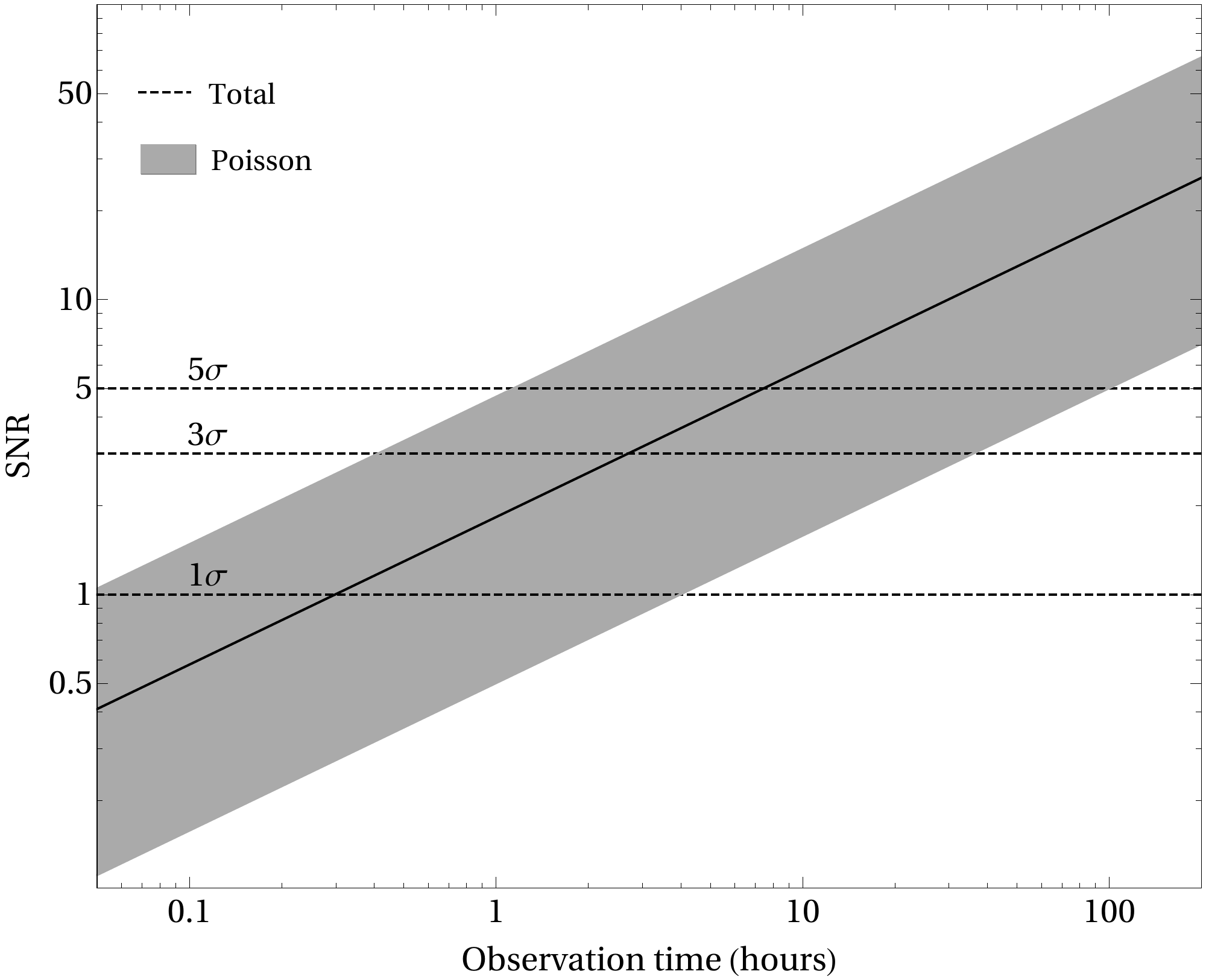}
\caption{\label{SNR-PBH-power}
    The SNR of LISA as a function of observation time for median total 
    SGWB (black curve) and associated uncertainties (grey shaded region),
    from the POBBHs (with a \textit{power-law} mass function) and BNSs.
    For BBHs, we adopt the best-fit value for $\al = 1.61$, 
    and the inferred local merger rate $R = 80\,^{+108}_{-56}$\, $\gpcyr$,
    which corresponds to $\fpbh = 3.8\,^{+2.3}_{-1.8} \times 10^{-3}$. 
    And for BNSs, we adopt $R = 1540_{-1220}^{+3200}$\,$\gpcyr$ 
    \citep{TheLIGOScientific:2017qsa}.
    The predicted median total background can be detected with 
    $\mathrm{SNR}=5$ after about $10$ hours of observation time.
    }
\end{figure}

\begin{table}[htbp!]
\begin{tabular}{c|c|c}
    &\ $\Omega_{\mathrm{GW}}(25 \, \mathrm{Hz})$ \
    &\ $\Omega_{\mathrm{GW}}(3 \times 10^{-3} \, \mathrm{Hz})$\,\\
	\hline
	BNS\, &  $0.7^{+1.5}_{-0.6} \times 10^{-9}$ 
          & $1.7^{+3.5}_{-1.4} \times 10^{-12}$ \\
	[.3em]
	\hline
	BBH\, & $1.8_{-1.3}^{+2.5} \times 10^{-9}$  
          & $4.3^{+5.9}_{-3.0} \times 10^{-12}$ \\
	[.3em]
	\hline
	Total\, & $2.5^{+4.0}_{-1.9} \times 10^{-9}$  
            & $6.0^{+9.4}_{-4.4} \times 10^{-12}$ \\
	[.2em]
\end{tabular}
    \caption{\label{Omegaf-PBH-power}
    Estimates of the background energy density $\ogw (\nu)$ at the most
    sensitive frequencies of LIGO (near $25$ Hz) and LISA 
    (near $3\times 10^{-3}$ Hz) for each of the BNS, POBBH
    (with a \textit{power-law} PDF) and 
    total background contributions, 
    along with the $90\%$ Poisson error bounds.
	}
\end{table}

\begin{figure}[htbp!]
\centering
\includegraphics[width = 0.48\textwidth]{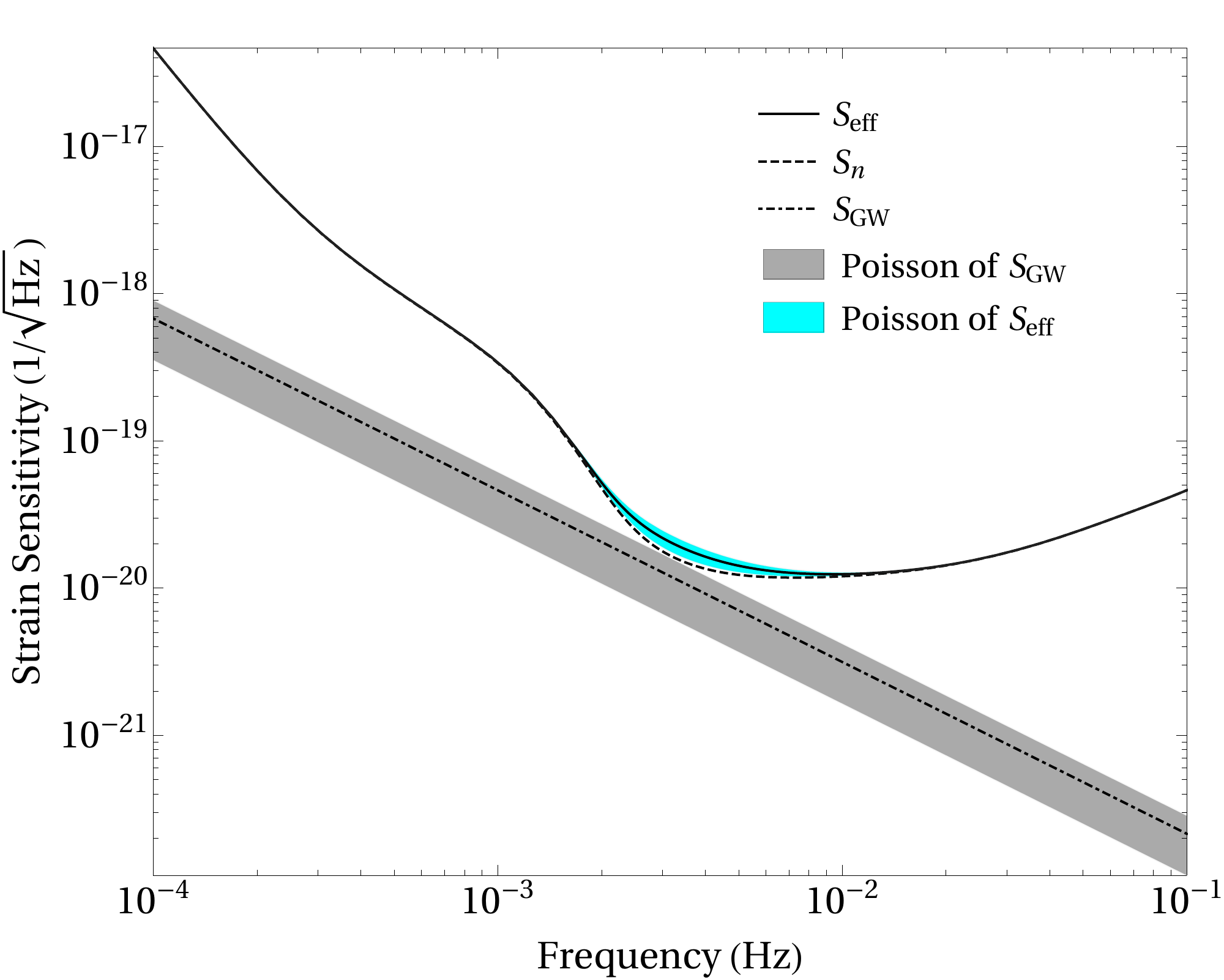}
\caption{\label{lisa-sensitivity-PBH-power}
		The effective strain sensitivity $S_{\mathrm{eff}}$ (black solid curve) of LISA
        and its Poisson uncertainties (cyan region),
        due to the effect of the total SGWB from POBBHs 
        (with a \textit{power-law} PDF) and BNSs.
        We also show LISA's strain sensitivity $S_n$ (dashed curve),
        and $S_{\mathrm{GW}}$ (dot-dashed curve) 
        along with its Poisson uncertainties (grey shaded region).
    }
\end{figure}

\begin{figure}[htbp!]
	\centering
	\includegraphics[width = 0.48\textwidth]{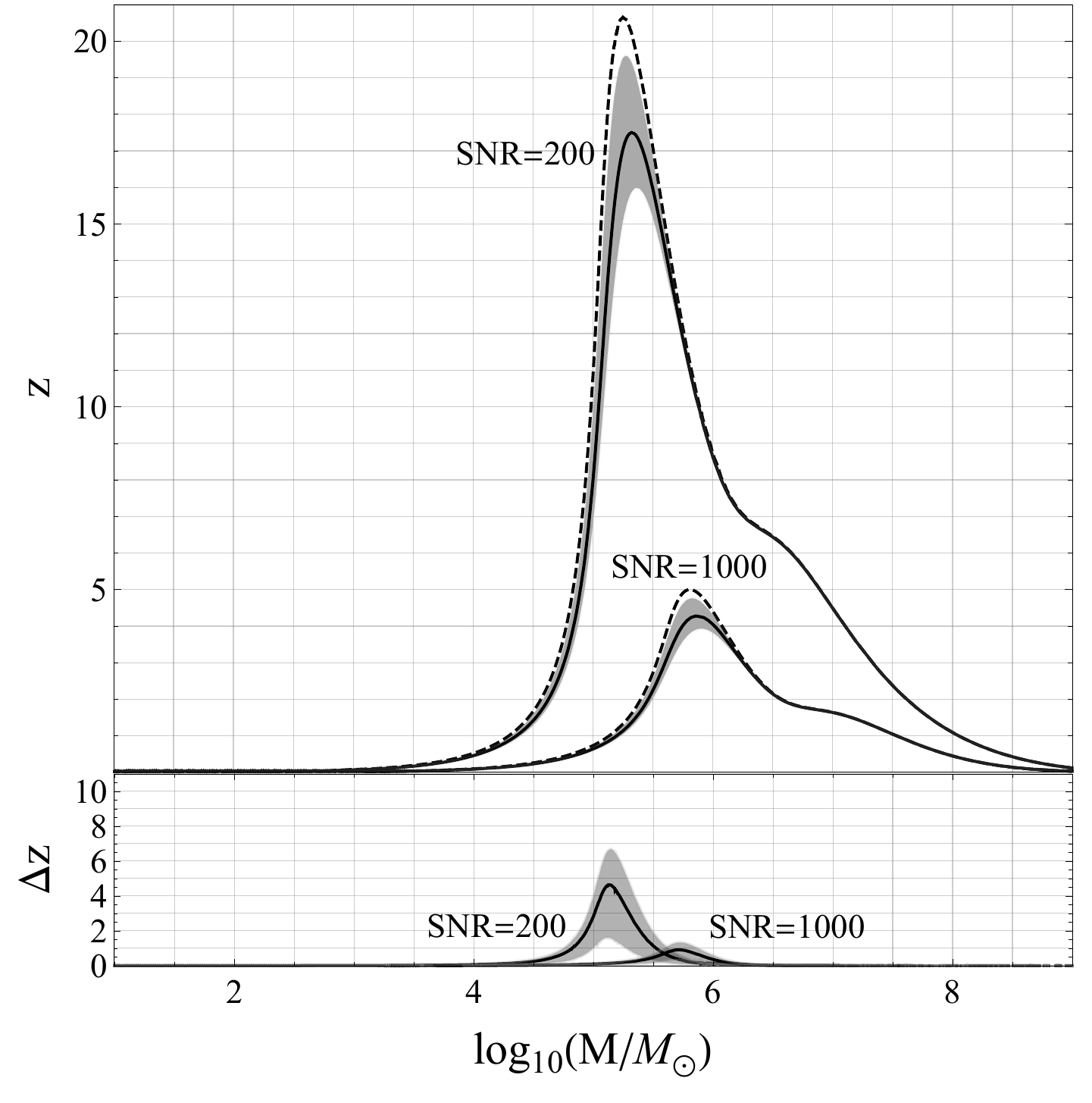}
	\caption{\label{z-PBH-power}
        The impacts of the total SGWB from POBBHs 
        (with a \textit{power-law} PDF) and BNSs, 
        on the largest detectable redshift $z$ of MBHB (with total mass $M$) coalescences for LISA.
        The mass ratio is set to $q=0.2$ following \cite{Audley:2017drz}. 
        The upper panel shows the contours of $\SNR=200$ and $\SNR=1000$ 
        for LISA (dashed curves), 
        together with the effect of SGWB (black curves) and the Poisson error 
        bars (grey shaded region). 
        The lower panel shows the residuals of corresponding contours. 
    }
\end{figure}

Utilizing \Eq{OmegaGW}, we then calculate the corresponding SGWB.
The result is shown in \Fig{OmegaGW-PBH-power}, 
indicating that the total SGWB from both POBBHs 
(with a \textit{power-law} PDF) and BNSs has a high possibility
to be detected by the future observing runs of \lvc\ and LISA.
In general, a variation of the PBH mass function will affect the profile, 
\eg the cutoff frequency and the magnitude,
of the energy spectrum $\ogw$.
To illustrate this impact, we plot a dotted line in \Fig{OmegaGW-PBH-power}, 
showing the background with $\Mmin = 0.5 \Msun$, 
by fixing $\{\al, R\}$ to their best-fit values as well.
The result indicates that the decreasing of $\Mmin$ will increase the 
population of the lighter PBHs and hence raise the cutoff frequency.
The enhancement of $\ogw$ is mainly due to the extra contribution from the
POBBHs with mass range $0.5 \sim 5 \Msun$.  
Note that LIGO's O2 result implies $\Mmin$ may not be too small; otherwise, SGWB will exceed the upper bround from LIGO's O2.

The energy spectra from both the POBBHs 
(with a \textit{power-law} PDF) and BNSs are well approximated by 
$\ogw \propto \nu^{2/3}$ at low frequencies covering both
LISA and LIGO's bands, 
where the dominant contribution is from the inspiral phase.
We also summarize the background energy densities $\ogw (\nu)$ 
at the most sensitive frequencies of LIGO (near $25$ Hz) and LISA 
(near $3\times 10^{-3}$ Hz) in \Table{Omegaf-PBH-power}.

\Fig{SNR-PBH-power} shows the expected accumulated SNR of LISA as a 
function of observing time.
The predicted median total background from POBBHs 
(with a \textit{power-law} PDF) and BNSs may be 
identified with $\SNR = 5$ after about $10$ hours of observation.
The total background could be identified with $\SNR = 5$ within $2$ hours 
of observation for the most optimistic case,  
and after about $5$ days for the most pessimistic case.
The strain sensitivity curves for LISA are shown in
\Fig{lisa-sensitivity-PBH-power}.
The effect on SNR of MBHB coalescences due to the unsolved SGWB signal 
is shown in \Fig{z-PBH-power}, 
indicating the precise measurement of the seeds of MBHs
above $10^5 \Msun$ in high formation redshift will be significantly affected 
by the confusion noise of the unsolved SGWB. 

\FloatBarrier
\subsection{Lognormal mass function}

We now consider another mass distribution, which has a lognormal form
\citep{Dolgov:1992pu},
\e\label{log}
 P(m) = \frac{1}{\sqrt{2 \pi} \s m} 
   \exp\(-\frac{\ln^2(m/m_c)}{2 \s^2}\),
\q
where $m_c$ and $\s$ give the peak mass of $m P(m)$ and the width of 
mass spectrum, respectively.
In this model, the free parameters are $\{\vth, R\} = \{m_c, \s, R\}$.

\begin{figure}[htbp!]
    \centering
    \includegraphics[width = 0.48\textwidth]{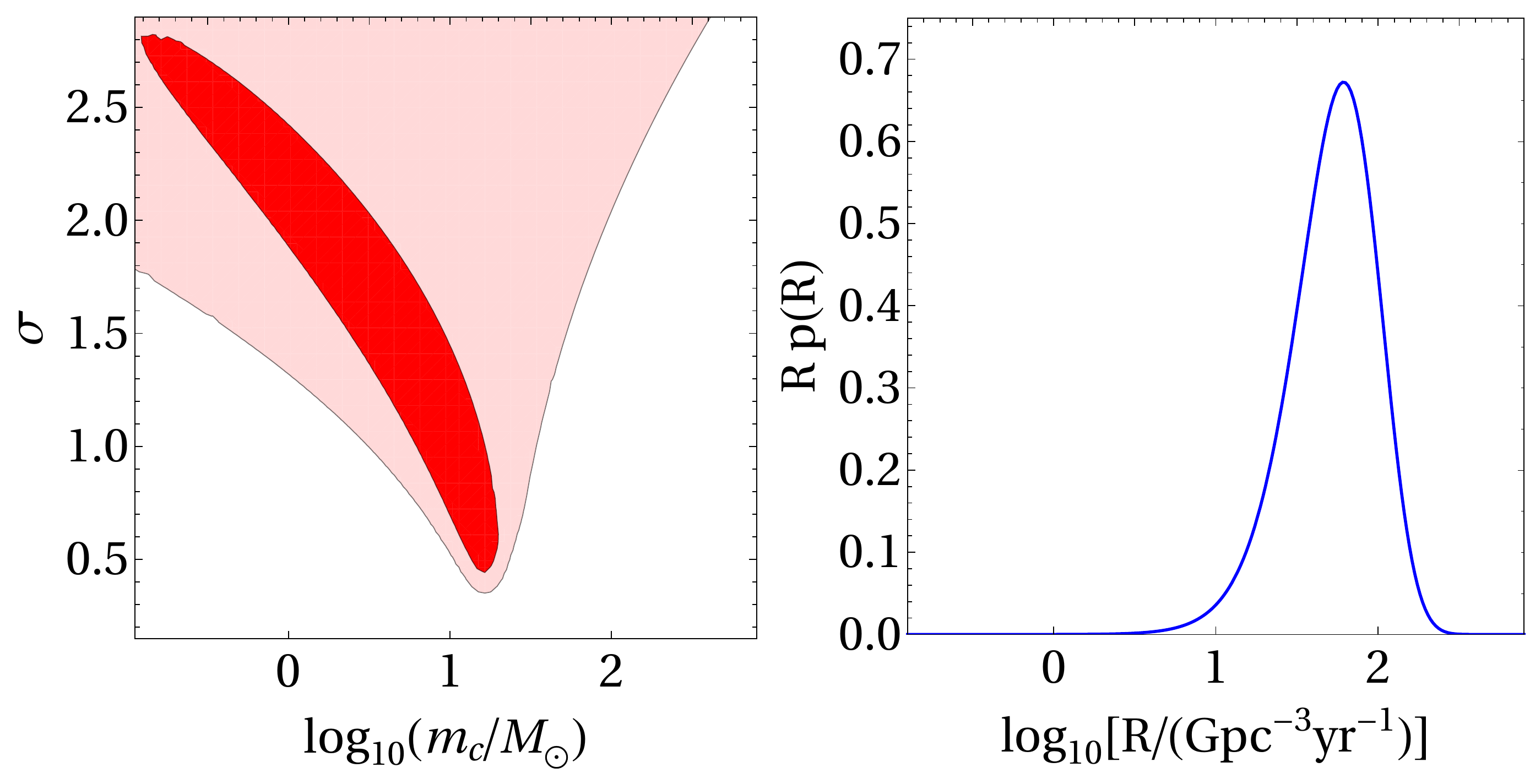}
    \caption{\label{posterior-PBH-log} 
    The posterior distributions for $\{\vth, R\} = \{m_c, \s, R\}$ of
    \textit{lognormal} mass function for PBHs, at the $68\%$ and $95\%$ 
    credible level, respectively, by using $3$ events from 
    LIGO's O1 observing run.    
    }
\end{figure}

\begin{figure}[htbp!]
\centering
\includegraphics[width = 0.48\textwidth]{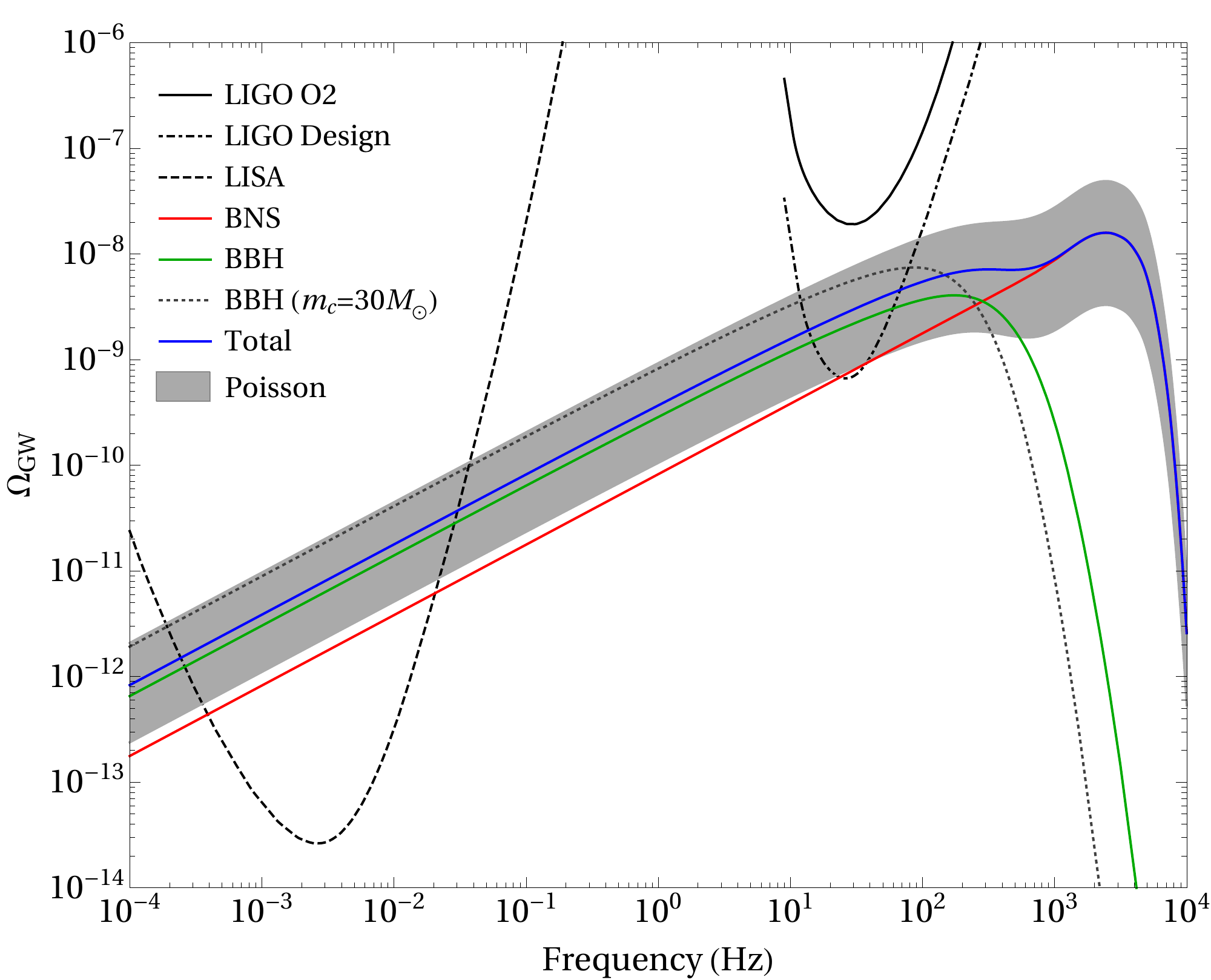}
\caption{\label{OmegaGW-PBH-log} 
    The predicted SGWB from the BNSs and POBBHs
    with a \textit{lognormal} mass function. 
    The red and green curves are backgrounds from the BNSs and BBHs, respectively. 
    The total (BNS and BBH) background is shown in the blue curve, while
    its Poisson error bars are in the grey shaded region.
    For BBHs, we adopt the best-fit values for 
    $\{m_c, \s\} = \{14.8\, \Msun,\,0.65\}$, 
    and the inferred local merger rate $R = 55\,^{+74}_{-38}$\, $\gpcyr$,
    which corresponds to $\fpbh = 2.8\,^{+1.6}_{-1.3} \times 10^{-3}$.
    And for BNSs, we adopt $R = 1540_{-1220}^{+3200}$\,$\gpcyr$ 
    \citep{TheLIGOScientific:2017qsa}.
    The dotted line shows the background from BBHs with $m_c = 30 \Msun$, 
    by fixing $\s = 0.65$ and $R = 55$\, $\gpcyr$.
    We also show the expected PI curves for LISA with $4$ years of 
    observation (dashed) and
    LIGO's observing runs of O2 (black) and design sensitivity (dot-dashed).
    The PI curves for LISA and LIGO's design sensitivity cross the Poisson
    error region, indicating the possibility to detect this background or
    set upper limits on the population parameters $\{m_c, \s, R\}$.
    }
\end{figure}

\begin{table}[htbp!]
\begin{tabular}{c|c|c}
    &\ $\Omega_{\mathrm{GW}}(25 \, \mathrm{Hz})$ \
    &\ $\Omega_{\mathrm{GW}}(3 \times 10^{-3} \, \mathrm{Hz})$\,\\
	\hline
	BNS\, &  $0.7^{+1.5}_{-0.6} \times 10^{-9}$ 
          & $1.7^{+3.5}_{-1.4} \times 10^{-12}$ \\
	[.3em]
	\hline
	BBH\, & $2.0^{+2.7}_{-1.4} \times 10^{-9}$  
          & $6.3^{+8.5}_{-4.2} \times 10^{-12}$ \\
	[.3em]
	\hline
	Total\, & $2.7^{+4.2}_{-2.0} \times 10^{-9}$  
            & $8.0^{+12}_{-5.6} \times 10^{-12}$ \\
	[.2em]
\end{tabular}
    \caption{\label{Omegaf-PBH-log}
    Estimates of the background energy density $\ogw (\nu)$ at the most
    sensitive frequencies of LIGO (near $25$ Hz) and LISA 
    (near $3\times 10^{-3}$ Hz) for each of the BNS, POBBH
    (with a \textit{lognormal} PDF) and 
    total background contributions, 
    along with the $90\%$ Poisson error bounds.
	}
\end{table}

\begin{figure}[htbp!]
    \centering
    \includegraphics[width = 0.48\textwidth]{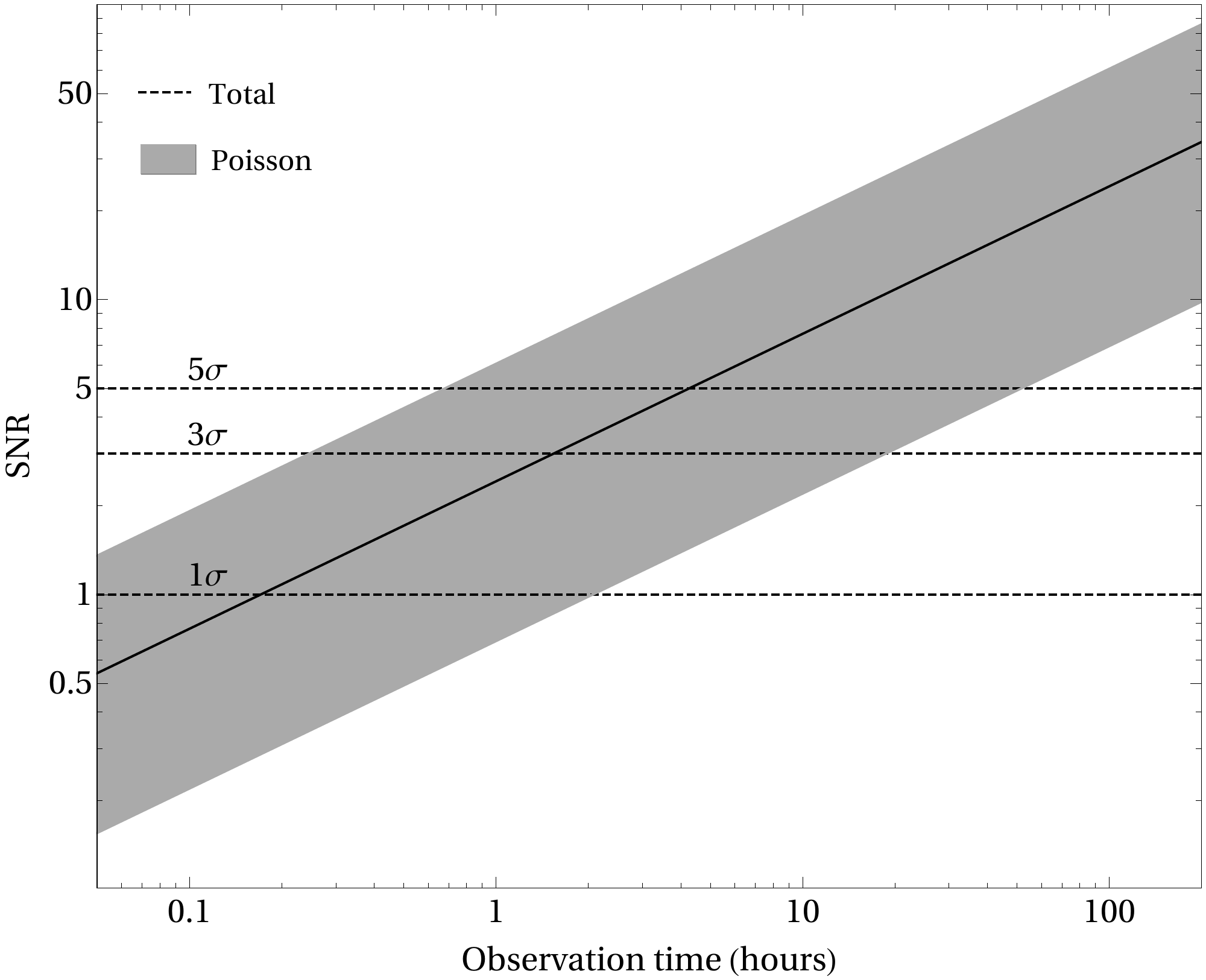}
    \caption{\label{SNR-PBH-log} 
    The SNR of LISA as a function of observation time for median total 
    SGWB (black curve) and associated uncertainties (grey shaded region),
    from POBBHs (with a \textit{lognormal} mass function) and BNSs.
    Here, we adopt the best-fit values for 
    $\{m_c, \s\} = \{14.8\, \Msun,\,0.65\}$, 
    and the inferred local merger rate $R = 55\,^{+74}_{-38}$\, $\gpcyr$,
    which corresponds to $\fpbh = 2.8\,^{+1.6}_{-1.3} \times 10^{-3}$.
    The predicted median total background can be detected with 
    $\mathrm{SNR}=5$ after about $5$ hours of observation time.
    }
\end{figure}

\begin{figure}[htbp!]
	\centering
	\includegraphics[width = 0.48\textwidth]{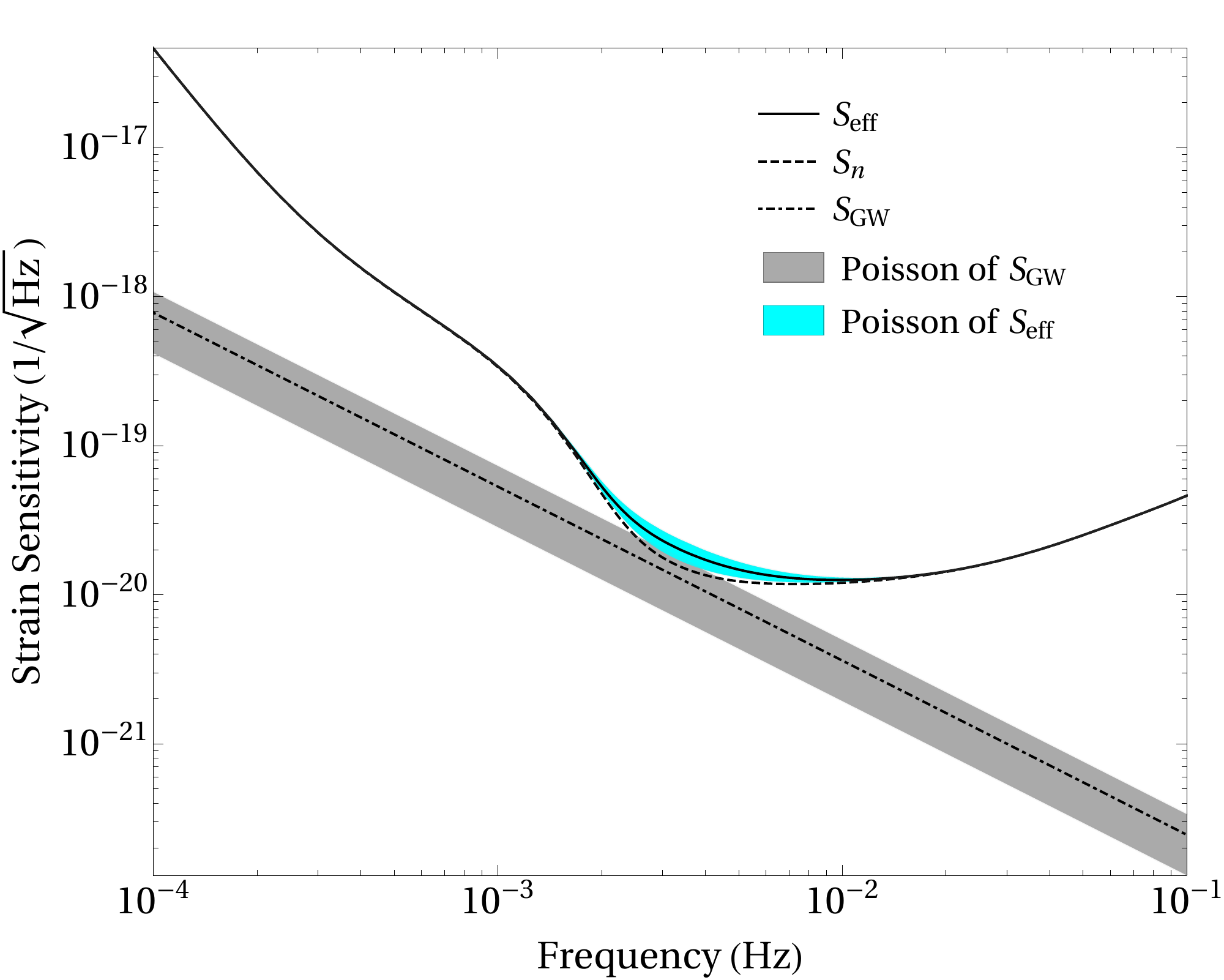}
	\caption{\label{lisa-sensitivity-PBH-log}
		The effective strain sensitivity $S_{\mathrm{eff}}$ (black solid curve) of LISA
        and its Poisson uncertainties (cyan region),
        due to the effect of the total SGWB from POBBHs 
        (with a \textit{log-normal} PDF) and BNSs.
        We also show LISA's strain sensitivity $S_n$ (dashed curve),
        and $S_{\mathrm{GW}}$ (dot-dashed curve) 
        along with its Poisson uncertainties (grey shaded region).
    }
\end{figure}

\begin{figure}[htbp!]
	\centering
	\includegraphics[width = 0.48\textwidth]{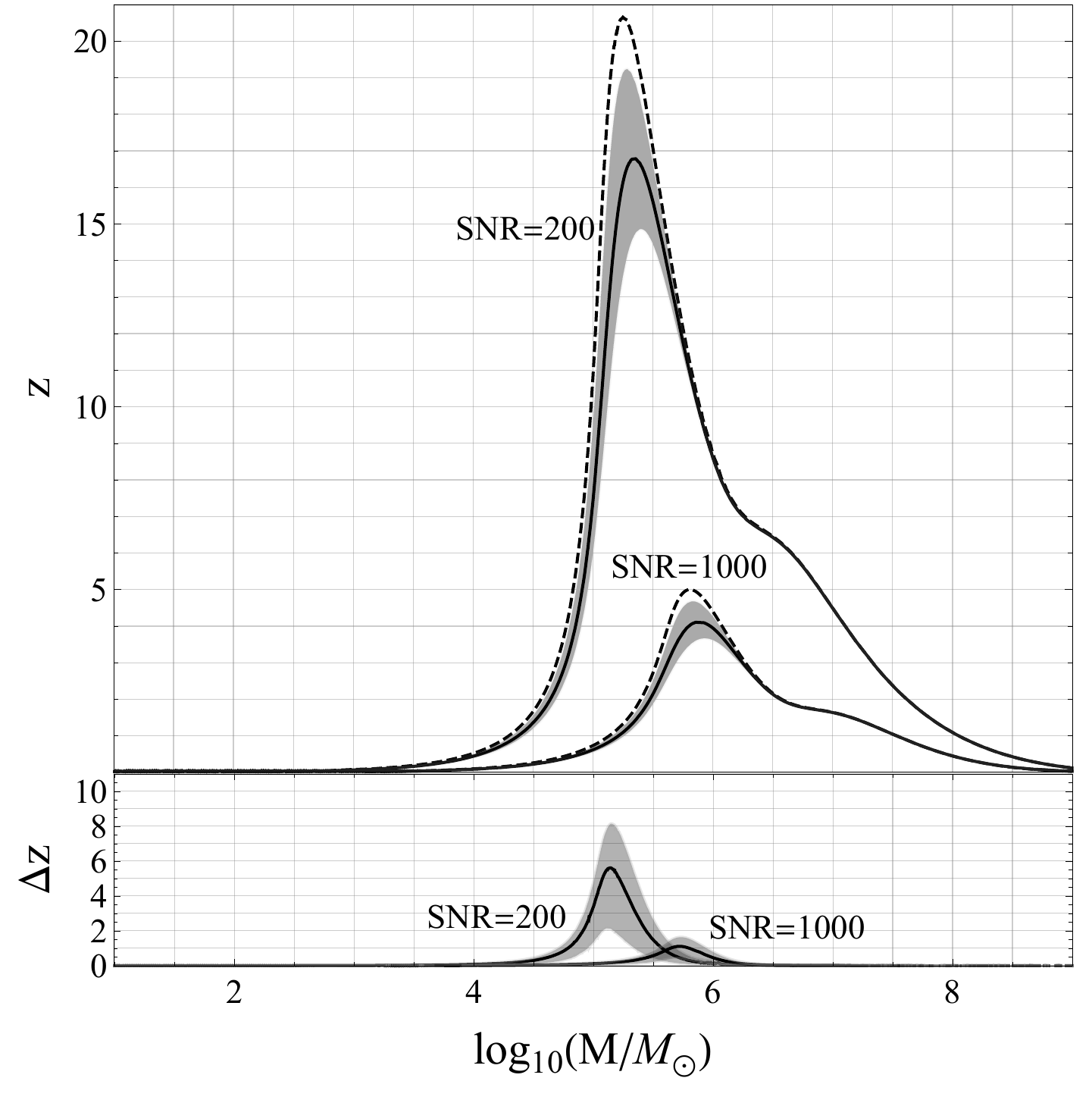}
	\caption{\label{z-PBH-log}
        The impacts of the total SGWB from POBBHs 
        (with a \textit{lognormal} PDF) and BNSs, 
        on the largest detectable redshift $z$ of MBHB (with total mass $M$) coalescences for LISA.
        The mass ratio is set to $q=0.2$ following \cite{Audley:2017drz}. 
        The upper panel shows the contours of $\SNR=200$ and $\SNR=1000$ 
        for LISA (dashed curves), 
        together with the effect of SGWB (black curves) and the Poisson error 
        bars (grey shaded region). 
        The lower panel shows the residuals of corresponding contours. 
    }
\end{figure}

Accounting for the selection effects and using $3$ events from LIGO's
O1 run, we find the best-fit results for $\vth$ are 
$\{m_c, \s\} = \{14.8\, \Msun,\, 0.65\}$.
Fixing $\vth$ to their best-fit values,
we obtain the median value and $90\%$ equal tailed credible
interval for the local merger rate,
$R = 55\,^{+74}_{-38}$\, $\gpcyr$.
From the posterior distribution of local merger rate $R$,
we then infer the fraction of PBHs in CDM 
$\fpbh = 2.8\,^{+1.6}_{-1.3} \times 10^{-3}$.
Such an abundance of PBHs is consistent with previous estimations that
$10^{-3} \lesssim \fpbh \lesssim 10^{-2}$,
confirming that the dominant fraction of CDM should not 
originate from the stellar mass PBHs
\citep{Sasaki:2016jop,Ali-Haimoud:2017rtz,Raidal:2017mfl,%
Kocsis:2017yty,Chen:2018czv}. 
The posterior distributions are shown in \Fig{posterior-PBH-log}. 
Compared to the results given in \cite{Raidal:2017mfl}, we see that, 
with the sensitivity of LIGO considered, a more restrictive constrains 
on the PDFs could be achieved.

Utilizing \Eq{OmegaGW}, we then calculate the corresponding SGWB.
The result is shown in \Fig{OmegaGW-PBH-log}, 
indicating that the total SGWB from both POBBHs 
(with a \textit{lognormal} PDF) and BNSs has a high possibility
to be detected by the future observing runs of \lvc\ and LISA.
To illustrate the impact of mass function on the profile of $\ogw$, 
we also plot a dotted line in \Fig{OmegaGW-PBH-log}, 
showing the background with $m_c = 30 \Msun$, 
by fixing $\{\s, R\}$ to their best-fit values.
The result indicates that the shifting of the central mass $m_c$ to a larger value
will decrease the cutoff frequency and increase the magnitude of $\ogw$, and vise versa. 

The SGWB for the \textit{lognormal} mass function was earlier calculated in \cite{Raidal:2017mfl} (see Fig.~2 therein).
They obtain a larger $\ogw$ than ours, indicating LIGO's O1 and O2 have the 
possibility to detect the SGWB from POBBHs.
There are a few reasons for this discrepancy.
Firstly, \cite{Raidal:2017mfl} inferred the parameters of the lognormal PDF 
by fitting the mass function instead of the merger rate distribution 
with LIGO's events and estimate the sensitivity
of LIGO by restricting the mass range to $7 \sim 50 \Msun$.
Here, we should note that the events of LIGO may not represent 
the intrinsic mass function due to the selection bias, 
of which the impact was ignored by \cite{Raidal:2017mfl}.
Their best-fits are $\{m_c, \s\} = \{33\, \Msun,\,0.8\}$, 
which is quite different from ours.
Secondly, \cite{Raidal:2017mfl} used the local merger rate $12 \sim 213$\, 
$\gpcyr$ derived from SOBBHs \citep{Abbott:2017vtc},
although which might serve as a good conservative estimation,
to infer the fraction of PBHs $\fpbh$.
In this paper, however, we improve their results by fitting the merger rate 
distribution using a full hierarchical Bayesian analysis,
and obtain $R = 55\,^{+74}_{-38}$\, $\gpcyr$ for POBBHs.

The energy spectra from both the POBBHs 
(with a \textit{lognormal} PDF) and BNSs are well approximated by 
$\ogw  \propto \nu^{2/3}$ at low frequencies covering both
LISA and LIGO's bands, 
where the dominant contribution is from the inspiral phase.
We also summarize the background energy densities $\ogw (\nu)$ 
at the most sensitive frequencies of LIGO (near $25$ Hz) and LISA 
(near $3\times 10^{-3}$ Hz) in \Table{Omegaf-PBH-log}.

\Fig{SNR-PBH-log} shows the expected accumulated SNR of LISA as a 
function of observing time.
The predicted median total background from POBBHs 
(with a \textit{lognormal} PDF) and BNSs may be 
identified with $\mathrm{SNR} = 5$ after about $5$ hours of observation.
The total background could be identified with $\mathrm{SNR} = 5$ 
within $1$ hours of observation for the most optimistic case,  
and after about $3$ days for the most pessimistic case.
The strain sensitivity curves for LISA are shown in
\Fig{lisa-sensitivity-PBH-log}.
The effect on SNR of MBHB coalescences due to unsolved SGWB signal 
is shown in \Fig{z-PBH-log}, 
indicating the precise measurement of the seeds of MBHs
above $10^5 \Msun$ in high formation redshift will be significantly affected 
by the confusion noise of the unsolved SGWB.

\section{\label{discuss}Summary and Discussion}
In this paper, we compute the total SGWB arising from both
the BBH and BNS mergers.
The influences of this SGWB on LISA's detection abilities is also investigated.
Two mechanisms for BBH formation, the astrophysical and primordial origins,
are considered separately.

For sBHs, we adopt the widely accepted ``Vangioni" model 
\citep{Dvorkin:2016wac}.
For the PBHs, we consider two popular but distinctive mass functions,
the power-law and lognormal PDFs, respectively.
For the power-law case, we infer the local merger rate to be
$R = 80\,^{+108}_{-56}$\, $\gpcyr$,
which corresponds to $\fpbh = 3.8\,^{+2.3}_{-1.8} \times 10^{-3}$;
while for the lognormal case, $R = 55\,^{+74}_{-38}$\, $\gpcyr$
and $\fpbh = 2.8\,^{+1.6}_{-1.3} \times 10^{-3}$. 
Comparing to the lognormal mass function, 
the power-law one implies a relatively lighter BBH mass distribution
and is compensated by a larger local merger rate, for consistency with
the event rate of \lvc.
Note that for both PDFs of PBHs, the inferred abundance of PBHs 
$\fpbh$ is consistent with previous estimations that
$10^{-3} \lesssim \fpbh \lesssim 10^{-2}$,
confirming that the dominant fraction of CDM should not 
originate from the stellar mass PBHs
\citep{Sasaki:2016jop,Ali-Haimoud:2017rtz,Raidal:2017mfl,%
Kocsis:2017yty,Chen:2018czv}. 

The resulting amplitude of SGWB from PBHs is significantly overall larger than
the previous estimation in \cite{Mandic:2016lcn}, 
which adopted the late Universe scenario
and assumed all the PBHs are of the same mass.
There are two reasons to account for this discrepancy.
One is that the early Universe scenario predicts much larger local
merger rate than the late Universe case
($R = 16\, \gpcyr$ in \cite{Mandic:2016lcn}).
Another one is that the merger rate (see \Eq{calR}) of the early Universe model
is strongly dependent on the redshift and sharply increase with redshift.
However, the merger rate of the late Universe model
is weakly dependent on redshift and slightly increases with redshift.
We refer to \cite{Mandic:2016lcn} for more details on the late Universe model.
We should emphasize that the above discussion applies only to the late Universe
scenario with a monochromatic mass function.
For the late Universe scenario with a general mass function,
the merger rate could be significantly enhanced
and the amplitude of SGWB could be greatly increased \citep{Clesse:2016ajp}.

Furthermore, PBHs contribute a stronger (at least comparable if 
we consider the uncertainties on the formation models of sBHs)
SGWB than that from the sBHs
(see Figs.~\ref{OmegaGW-sBH}, \ref{OmegaGW-PBH-power}, \ref{OmegaGW-PBH-log},
and also Tables~\ref{Omegaf-sBH}, \ref{Omegaf-PBH-power}, \ref{Omegaf-PBH-log}).
This is due to that the merger rate densities from PBHs and sBHs have 
quite different dependences on the BH masses and redshift.
Especially, the merger rate of PBHs sharply increases with redshift;
while the merger rate of sBHs first increases, 
then peaks around $z \sim 1-2$, and last rapidly decreases with redshift. 

In addition, the background energy densities from primordial 
and astrophysical BBH mechanisms, 
both show no clear deviation from the power law spectrum 
$\ogw \propto \nu^{2/3}$,
within \lvc\ and LISA sensitivity band. 
Thanks to their similar effects on the spectra, 
distinguishing the backgrounds between POBBHs (the early Universe scenario)
and SOBBHs will be challenging.
However, \cite{Clesse:2016ajp} claimed that the SGWB of POBBHs from the 
late Universe could potentially deviate $\ogw \propto \nu^{2/3}$ at the pulsar
timing arrays (PTA) frequencies and even at the frequencies higher enough to 
be probed by LISA.
The feature presented in \cite{Clesse:2016ajp} may be used to distinguish different formation channels of BBHs.

Finally, the total SGWB from both BBHs (whether astrophysical or primordial
origin) and BNSs has a high possibility to be detected by the future 
observing runs of \lvc\, and LISA,
as could be seen from 
Figs.~\ref{OmegaGW-sBH}, \ref{SNR-sBH},
\ref{OmegaGW-PBH-power}, \ref{SNR-PBH-power},
\ref{OmegaGW-PBH-log}, \ref{SNR-PBH-log}.
This SGWB also contributes an additional source of confusion noise to LISA's
total noise curve (see Figs.~\ref{lisa-sensitivity-sBH},
\ref{lisa-sensitivity-PBH-power}, \ref{lisa-sensitivity-PBH-log}), 
and hence weakens LISA's detection abilities.
For instance, the detection of MBHB coalescences
is one of the key missions of LISA, 
and the largest detectable redshift of MBHB mergers can be significantly 
reduced (see Figs.~\ref{z-sBH}, \ref{z-PBH-power}, \ref{z-PBH-log}).
Therefore, further analysis is needed to subtract the SGWB signals
from the data in order to improve the performance of the detectors.

\acknowledgments
We thank the anonymous referee for valuable suggestions and comments.
We would also like to thank Yun-Kau Lau, Zheng-Cheng Liang, Lang Liu, Hao Wei, 
Wen Zhao, Yuetong Zhao and Xiao-Bo Zou  for useful conversations. 
We acknowledge the use of HPC Cluster of ITP-CAS. 
This work is supported by grants from NSFC 
(grant No. 11335012, 11575271, 11690021, 11747601), 
the Strategic Priority Research Program of Chinese Academy of Sciences 
(Grant No. XDB23000000), Top-Notch Young Talents Program of China, 
and Key Research Program of Frontier Sciences of CAS. 
This research has made use of data, software and/or web tools obtained 
from the Gravitational Wave Open Science Center (https://www.gw-openscience.org), 
a service of LIGO Laboratory, the LIGO Scientific Collaboration and the Virgo
Collaboration. LIGO is funded by the U.S. National Science Foundation. 
Virgo is funded by the French Centre National de Recherche Scientifique (CNRS), 
the Italian Istituto Nazionale della Fisica Nucleare (INFN) and the Dutch Nikhef,
with contributions by Polish and Hungarian institutes.

\bibliographystyle{apj}
\bibliography{./bibfile}

\end{document}